\documentclass[hyperref={colorlinks,citecolor=red,urlcolor=black,bookmarks=false,hypertexnames=true}]{article}

\usepackage{arxiv}

\usepackage[utf8]{inputenc}
\usepackage[T1]{fontenc}
\usepackage{lmodern}
\usepackage{url}
\usepackage{booktabs}
\usepackage{amsfonts}
\usepackage{nicefrac}
\usepackage{microtype}
\usepackage{graphicx}
\usepackage{subcaption}
\usepackage{doi}

\usepackage{colortbl}
\usepackage{listings}
\usepackage{bm}
\usepackage{physics}
\usepackage{amssymb}
\usepackage{longtable}
\usepackage{amsmath,accents}
\usepackage{cleveref}

\usepackage[ruled,vlined]{algorithm2e}
\usepackage{appendix}
\usepackage{cite}
\usepackage{multirow}
\usepackage{array}

\newcommand{\sinc}{\mathop{\mathrm{sinc}}}

\usepackage{chngcntr}
\usepackage{epigraph}
\usepackage{gensymb}
\usepackage{import}
\usepackage{xifthen}
\usepackage{pdfpages}
\usepackage{transparent}

\SetKwComment{Comment}{/* }{ */}

\newcommand{\ttl}{A Two-Mirror Faceted Projection System\\ for EUV Lithography}

\title{\ttl}

\date{}

\author{ \href{https://orcid.org/0000-0002-4930-1846}{\includegraphics[scale=0.06]{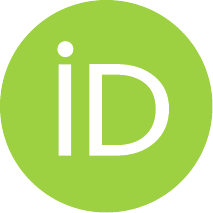}\hspace{1mm}Vasiliy A. Es'kin}\thanks{Corresponding author: Vasiliy Alekseevich Es'kin (\href{mailto:vasiliy.eskin@gmail.com}{vasiliy.eskin@gmail.com})}, {Egor V. Ivanov}, \href{https://orcid.org/0000-0001-8856-0114}{\includegraphics[scale=0.06]{orcid.pdf}\hspace{1mm}Olga V. Martynova} \\
	Department of Radiophysics, University of Nizhny Novgorod\\
	23 Gagarin Ave., Nizhny Novgorod 603022, Russia\\
	\href{mailto:vasiliy.eskin@gmail.com}{\texttt{vasiliy.eskin@gmail.com}}, {\texttt{iev90078@gmail.com}, \texttt{ruvin@list.ru}}
}

\renewcommand{\headeright}{}
\renewcommand{\undertitle}{}
\renewcommand{\shorttitle}{A Two-Mirror Faceted Projection System for EUV Lithography}

\renewcommand{\vec}{\mathbf}

\newcommand{\eps}{\varepsilon}

\usepackage{tikz}
\newcommand\tikznode[3][]%
{\tikz[remember picture,baseline=(#2.base)]
	\node[minimum size=0pt,inner sep=0pt,#1](#2){#3};%
}

\hypersetup{
	pdftitle={A Two-Mirror Faceted Projection System for EUV Lithography},
	pdfsubject={physics.optics,physics.comp-ph,physics.class-ph, physics.app-ph,cs.LG},
	pdfauthor={Vasiliy A. Es'kin, Egor V. Ivanov, Olga V. Martynova},
	pdfkeywords={Computational physics, Extreme Ultraviolet Lithography, Two-mirror projection system, High numerical aperture, Multilayer Bragg mirrors, Inverse Lithography Technology, Waveguide method, RCWA, Machine Learning}
}

\begin{document}
	\maketitle
	\vspace{-12mm}
	\begin{figure}[ht!]\centering
		\includegraphics[width=0.3\textwidth]{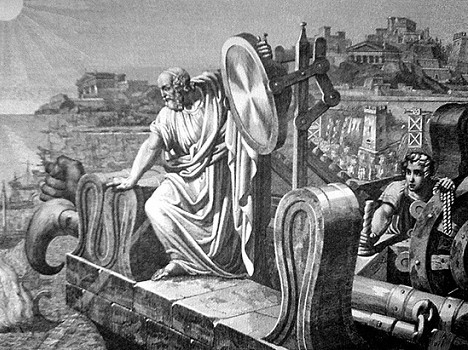}    
	\end{figure}
	{
		\setlength{\epigraphwidth}{0.5\textwidth}
		\epigraph{Happiness for everybody, free, and no one will go away unsatisfied!}{Arkady and Boris Strugatsky, \textit{Roadside Picnic}}
	}
	
	\begin{abstract}
		We propose an all-reflective two-mirror projection system for extreme ultraviolet (EUV) lithography operating at exposure wavelengths of $13.5$~nm (Mo/Si) and $11.2$~nm (Ru/Be), delivering a fourfold ($4\times$) demagnification of the periodic mask pattern at a numerical aperture approaching unity ($\mathrm{NA}_{\max} \approx 0.993$). In contrast to conventional EUV projection objectives that incorporate 6--10 aspheric mirrors with an overall optical throughput of less than $15\%$, the proposed design redirects each accepted discrete spatial diffraction order scattered by the mask onto the wafer via a dedicated pair of planar mirror facets. The number of reflections is strictly fixed at two for all accepted orders, retaining $50$--$60\%$ of the power leaving the mask in each accepted order. We derive a spatial geometry providing rigorous optical path length equalization across all diffraction orders, thereby removing order-dependent propagation phase shifts. Individually optimized 30-bilayer Bragg multilayer coatings are designed for each facet using the transfer matrix method combined with global evolutionary optimization algorithms. The architecture is generalized to a three-dimensional vector formulation with a two-dimensionally periodic mask. Utilizing inverse lithography technology, Fourier parameterization, and a differentiable electromagnetic modal waveguide solver, we solve the synthesis problem for binary absorber masks (La absorber on a Ru/Be/Sr multilayer mirror). We demonstrate simulated aerial images of sub-10-nm features on the wafer (isolated peaks with a full width at half maximum (FWHM) of approximately $5.4$~nm and line pairs with a critical dimension of $6$~nm) and find that the two peaks remain resolved for the tested wafer defocus values from $0$ to $5$~nm along the $z$-axis.
	\end{abstract}
	
	\keywords{Extreme Ultraviolet Lithography \and Two-mirror projection system \and High numerical aperture \and Multilayer Bragg mirrors \and Inverse Lithography Technology \and Waveguide method \and RCWA \and Computational physics \and Machine Learning}

	\section{Introduction}\label{intro}
	
	The modern technological world critically depends on semiconductor microchips, which lie at the heart of smartphones and personal computers, control automobiles and industrial robotics, route global internet traffic, and serve as the physical foundation for the ongoing wave of artificial intelligence systems. Global demand for computation, telecommunications, and machine perception is expanding at an explosive pace, and microelectronics has met this challenge for over half a century by doubling the number of active components on a chip approximately every two years in accordance with Moore's Law~\cite{VanSchoot2025}. At the foundation of this exponential progress lies the continuous downscaling of physical device dimensions, directly tied to the miniaturization of the individual transistor: smaller transistors yield higher integration density, superior operational speed, and enhanced energy efficiency. Optical photolithography serves as the primary engine of this miniaturization. It transfers the topology of the integrated circuit layout from a photomask onto a silicon wafer, thereby dictating the minimum printable feature size. The critical role of this tool in the economics of semiconductor manufacturing is reflected in its capital cost: while a 193-nm immersion lithography scanner costs on the order of 50~million euros, a state-of-the-art EUV scanner exceeds 300~million euros~\cite{Chkhalo2024}.
	
	The minimum feature size printable via photolithography is fundamentally constrained by physical laws and obeys Rayleigh's resolution equation, $\mathrm{CD} = k_1 \lambda/\mathrm{NA}$, where $\mathrm{CD}$ is the critical dimension, $k_1$ is a process-dependent factor (with a theoretical physical limit of $0.25$), $\lambda$ is the exposure wavelength, and $\mathrm{NA}$ is the numerical aperture of the projection optics~\cite{VanSchoot2025}. As is evident from this fundamental relationship, shrinking feature dimensions requires either reducing the exposure wavelength or increasing the numerical aperture. Over six decades, the exposure wavelength transitioned from 436~nm (mercury g-line) through 365, 248, and 193~nm down to 13.5~nm, while the numerical aperture in liquid immersion systems surpassed unity to reach 1.35~\cite{VanSchoot2025,IRDS2024}. Along this developmental trajectory, lithographers adopted advanced techniques, such as off-axis illumination (OAI) and multiple patterning. In 2018, EUV lithography became commercially viable for high-volume manufacturing following the introduction of reliable 13.5-nm plasma sources delivering output powers exceeding 125~W~\cite{VanSchoot2025}.
	
	The operational wavelength of 13.5~nm was selected because periodic Mo/Si Bragg multilayer coatings composed of 40--50 bilayers exhibit record near-normal reflectivities of $\sim 70\%$ per mirror surface~\cite{Chkhalo17}, and commercial scanner sources now deliver powers exceeding 400~W~\cite{VanDeKerkhof2022}. Standard NXE scanners operating at $\mathrm{NA} = 0.33$ print a minimum half-pitch of 13~nm, serving as the core workhorses of high-volume semiconductor manufacturing. The subsequent generation of High-NA EXE:5000 systems, with $\mathrm{NA} = 0.55$, pushes the resolution down to 8~nm. However, this is achieved via an anamorphic projection architecture ($4\,\times\,/\,8\,\times$) in which the depth of focus (DoF) is reduced by a factor of 2.94, the exposure field size is halved, and mirror angles of incidence are constrained by the narrow angular reflectance bandwidth of Mo/Si coatings ($\pm 11^{\circ}$)~\cite{VanSchoot2017,VanSchoot2025,Kalden2025,VanSetten2017}. Further increasing the numerical aperture to $\mathrm{NA} = 0.75$ and beyond, as required for printing critical dimensions below 6~nm, will necessitate an even greater number of mirrors within the all-reflective projection channel~\cite{Kalden2025,Shintake2024}.
	
	Each reflection in the illumination and projection optics absorbs a substantial fraction of the incident EUV power~\cite{Chkhalo2024}. This motivates alternative wavelengths, including $11.2$~nm with Ru/Be coatings and Xe-discharge sources~\cite{Chkhalo2024,Abramov2025}, short-wavelength multilayers in the 9--12~nm band~\cite{Shaposhnikov22,Polkovnikov_2020,Chkhalo17}, and architectures with fewer reflections. Shintake~\cite{Shintake2024} compares a conventional system with four illumination and six projection reflections with a proposed system having two illumination and two projection reflections. At an assumed reflectance of $0.65$ per surface, the corresponding transmissions are $0.65^{10}\approx1.3\%$ and $0.65^4\approx18\%$. The associated estimate of a reduction from 1~MW to approximately 80~kW concerns electricity used for EUV generation, not the total scanner consumption. A later four-mirror in-line design~\cite{Shintake2025} uses repeated encounters with its physical mirrors, giving eight reflections for $\mathrm{NA}=0.5$ and twelve for $\mathrm{NA}=0.7$. Mirror count must therefore be distinguished from reflection count.
	
	The design of the reflective optical systems described above relies, in most cases, on geometric ray-tracing methods, which represent crude approximations of electromagnetic wave propagation that neglect wave phenomena such as interference and diffraction. Incorporating diffractive elements directly into the projection optics has been explored previously. For instance, in a three-mirror all-reflective scheme operating at 13~nm, aspheric mirrors were replaced by reflection diffraction gratings on spherical substrates~\cite{Fukuda1995}, and later this line of research was extended using reflective Fresnel zone plates~\cite{Zheng2008,Zheng2011}. However, in those systems the diffractive structure merely reproduces the phase profile of an aspheric mirror, the total number of reflections within the channel is not reduced, and stringent monochromaticity requirements are imposed on the source. Separately, maskless EUV lithography concepts based on arrays of micro-electromechanical mirrors have been proposed as alternatives to physical masks~\cite{Choksi1999}, addressing mask cost, defect, and pellicle challenges, though requiring their own specialized projection architectures.
	
	In this work, we develop an alternative concept that pushes the reduction of reflective surfaces to its logical limit. We propose a two-mirror all-reflective projection system in which each spatial diffraction order scattered by the mask is redirected onto the wafer individually by a dedicated pair of planar mirror facets. The number of reflections per accepted channel between the mask and wafer is fixed at two, independent of the numerical aperture. Thus, values of $\mathrm{NA} \sim 1$ are fundamentally attainable alongside a fourfold ($4\times$) demagnification of the mask pattern. We consider both two-dimensional and three-dimensional problem formulations. In the 3D case, the mask represents a two-dimensionally periodic structure (a square grating), and each propagating diffraction order is mapped to its own corresponding facet pair across both mirrors. The analysis is performed in an approximation close to geometrical optics (GO), where propagating spatial diffraction orders are described as collimated paraxial beams with uniform cross-sectional amplitude.
	
	The paper is structured as follows. In Section~\ref{problem}, the electromagnetic diffraction problem for an EUV mask is formulated and the basic relations governing the two-mirror projection system are derived. Section~\ref{2dmirrors} presents the two-dimensional two-mirror projection system, including the iterative design of the mirror geometry, the optimization of individually tuned Bragg multilayer coatings for each facet, and the synthesis of binary absorber masks for a target aerial image using inverse lithography technology with Fourier-parameterized projection. Section~\ref{3Dmirrors} extends the formulation to three dimensions, generalizing the design to a two-dimensionally periodic mask with a square lattice and demonstrating 3D mask synthesis. Section~\ref{discussion} discusses the advantages and disadvantages of the proposed projection system. Finally, in Section~\ref{conclusion} concluding remarks are given. Appendix~\ref{appA} provides a rigorous analysis of the validity of the geometrical-optics beam approximation for centimeter-scale EUV beams.

	\section{Formulation of the Problem}\label{problem}
	
	We begin by examining the electromagnetic diffraction problem for an EUV mask in a two-dimensional formulation, subsequently extending the analysis to the structural design of a projection system providing a $4\times$ demagnification of the mask pattern onto the wafer.
	
\begin{figure}[ht!]\centering
		\includegraphics[width=0.7\textwidth]{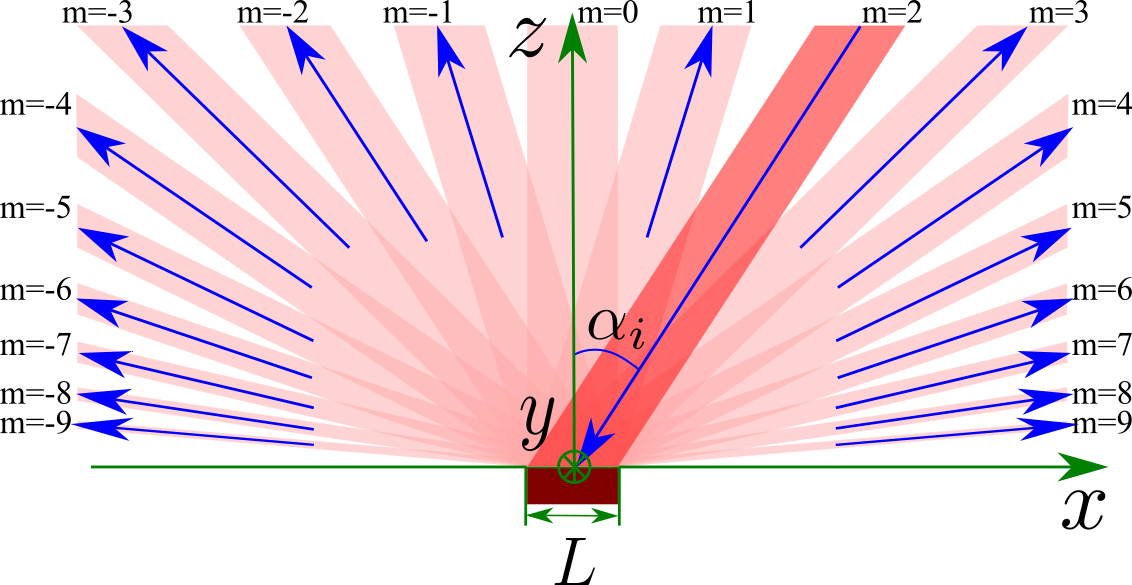}
		\caption{Electromagnetic beam scattering by a lithographic mask (schematic in the geometrical-optics approximation).}\label{fig1}
	\end{figure}

\begin{figure}[ht!]\centering
		\includegraphics[width=0.45\textwidth]{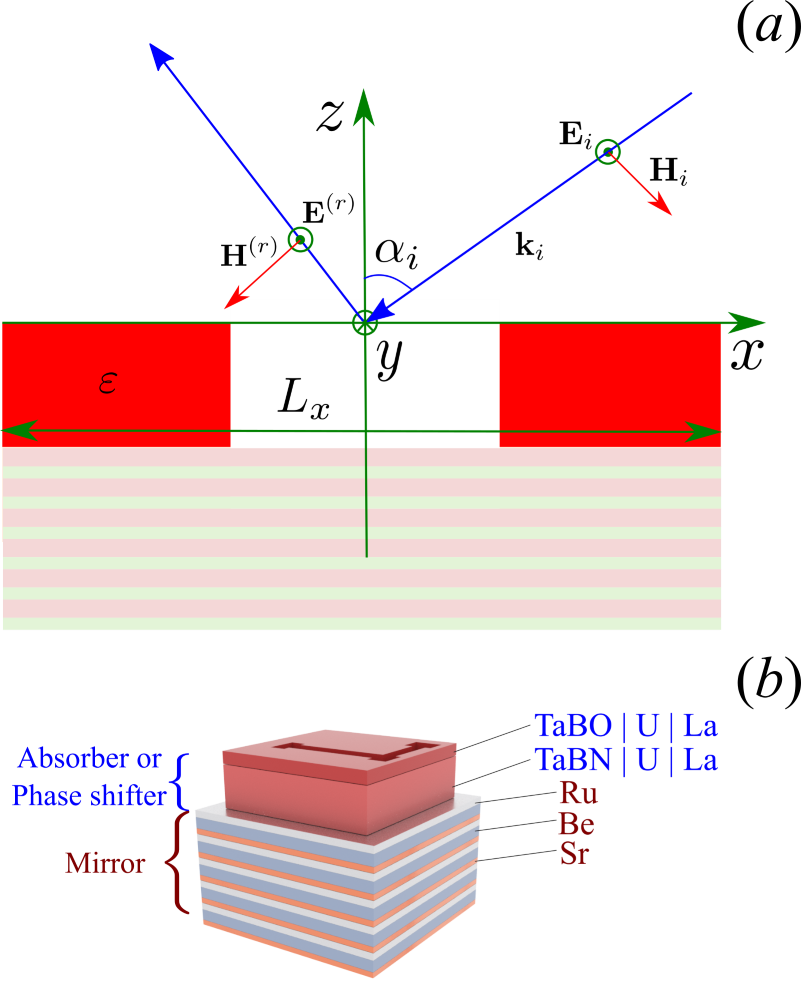}
		\caption{Geometry of the problem: (a) cross-section at $y = 0$ with mask spatial period $L_x$, (b) perspective view of the mask stack.}\label{fig1_b}
	\end{figure}

	The layered mask, consisting of $J$ layers, occupies the domain $[-D, 0]$ along the $z$-axis and $[-L/2, L/2]$ along the $x$-axis in a Cartesian coordinate system $(x, y, z)$, as illustrated in Fig.~\ref{fig1}, and is periodic along the $x$-direction with spatial period $L_x$. Each layer $j$ is homogeneous along the $z$-direction and has a complex permittivity $\eps_j = \eps_j(x)$. Optical constants are obtained from tabulated reference data~\cite{HENKE1993181,CenterXRayOpt}. Within the mask stack, one distinguishes patterned absorber (or phase-shifting) layers that define the composition of the scattered spectrum, and the underlying multilayer Bragg mirror substrate (see Fig.~\ref{fig1_b}). Here, the multilayer substrate is assumed to consist of periodic three-component Ru/Be/Sr stacks~\cite{eskin2026physicsinformedneuralsystemssimulation} or two-component Ru/Be (or Mo/Si) bilayers~\cite{Shaposhnikov22,Polkovnikov_2020,Chkhalo17}.
	
	The primary objective is to construct an all-reflective projection system that provides a $4\times$ demagnification of the mask pattern when projected onto the wafer. Under $4\times$ reduction, the spatial period on the wafer, $L_x^{(\mathrm{w})}$, satisfies $L_x^{(\mathrm{w})} = L_x/4$ (i.e., $L_x = 4 L_x^{(\mathrm{w})}$), where both the wafer plane and the mask plane are assumed to be parallel to the $xOy$ plane. Pattern transfer from the mask to the wafer is mediated by the electromagnetic field scattered by the mask and focused onto the wafer.
	
	Assuming the lateral mask dimension satisfies $L \gg L_x$, edge effects can be neglected, allowing the mask to be treated as an infinite periodic structure during the diffraction analysis. A TE-polarized monochromatic plane wave with angular frequency $\omega$ is obliquely incident on the mask. In the geometrical-optics (GO) beam approximation, the field amplitude is uniform within the beam envelope and zero elsewhere. The incident electric field is given, with $\exp(\mathrm{i}\omega t)$ time dependence dropped, by $\mathbf{E}^{(i)} = \mathbf{y}_0 E_0 \exp[-\mathrm{i}(\tilde{m}\kappa_x x - k_{z}^{(i)}z)]$, where $E_0$ is the electric field amplitude, $\tilde{m}\kappa_x$ and $k_{z}^{(i)}$ are the components of the wave vector ${\vec k}_0$ in free space ($k_0 = \left[\tilde{m}^2 \kappa_x^2 + \left(k_{z}^{(i)}\right)^2\right]^{1/2}$, $\kappa_x = 2\pi/L_x$, $k_0 = \omega/c$, where $c$ is the speed of light in free space), and superscript $(i)$ denotes the incident wave. The resulting boundary-value diffraction problem is solved using modal numerical methods~\cite{tanabe2024accelerating,yuan1991modeling,lucas1996efficient,Medvedev2023,EskinDD2025,eskin2026physicsinformedneuralsystemssimulation}. The scattered electric field in the vicinity of the mask is expanded in a series:
	\begin{equation}
		E_y^{(r)} = \sum_{m=-\infty}^{\infty} A_m^{(r)} \exp(-\mathrm{i}\kappa_x m x - \mathrm{i} k_{z;m}z),
		\label{eq:refl}
	\end{equation}
	where $A_m^{(r)}$ is the complex amplitude of the $m$-th scattered spatial harmonic, and $k_{z;m} = \left(k_0^2 - \kappa_x^2 m^2\right)^{1/2}$ (choosing the branch $\mathrm{Im}\, k_{z;m} \le 0$).
	
	For the projection-system design, only propagating diffraction orders (purely real $k_{z;m}$) are relevant. Thus, the scattered field of interest outside the near-field region of the mask is:
	\begin{equation}
		E_y^{(r)} = \sum_{m=-M}^{M} \tilde{A}_m^{(r)}(x, z) \exp(-\mathrm{i}\kappa_x m x - \mathrm{i} k_{z;m}z),
		\label{eq2}
	\end{equation}
	where $\tilde{A}_m^{(r)}(x, z) = A_m^{(r)}$ inside the spatial beam corresponding to order $m$ (see Fig.~\ref{fig1}), and $\tilde{A}_m^{(r)}(x, z) = 0$ outside it. The maximum propagating order index is $M = \lfloor k_0 / \kappa_x \rfloor$.
	
	To achieve a $4\times$ demagnified aerial image on the wafer, the propagating field spectrum at the wafer boundary must take the form:
	\begin{equation}
		E_y^{(\mathrm{w})} = \sum_{n=-N}^{N} B_n \exp[-\mathrm{i}\kappa_x^{(\mathrm{w})} n x - \mathrm{i} s_{\mathrm{w}} k_{z;n}^{(\mathrm{w})}\zeta_{\mathrm{w}}],
		\label{eq3}
	\end{equation}
	where $\zeta_{\mathrm{w}}=z-Z_{\mathrm{w}}$ is measured from the wafer plane, $s_{\mathrm{w}}=+1$ in Examples 1, 3, and 4 and $s_{\mathrm{w}}=-1$ in Example 2, and $B_n$ is the complex amplitude of the $n$-th harmonic at $\zeta_{\mathrm{w}}=0$, $\kappa_x^{(\mathrm{w})} = 2\pi / L_x^{(\mathrm{w})}$, and $k_{z;n}^{(\mathrm{w})} = \left[k_0^2 - (\kappa_x^{(\mathrm{w})})^2 n^2\right]^{1/2}$. Under $4\times$ reduction, the maximum propagating order transmitted to the wafer is $N = \lfloor M/4 \rfloor$ (ideally $M = 4N$; the difference arises from the cutoff of orders $n > N$ as evanescent waves).
	
\begin{figure}[ht!]\centering
		\includegraphics[width=0.45\textwidth]{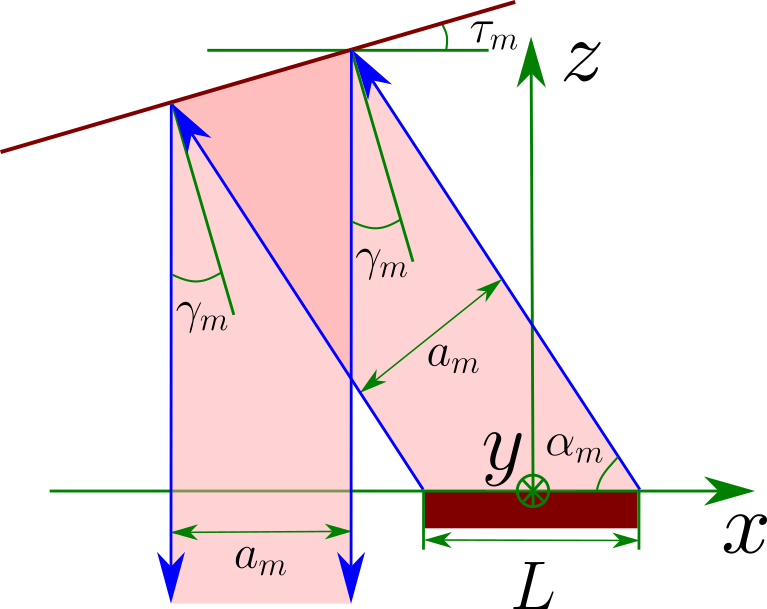}
		\caption{Reflection of the $m$-th diffraction beam from a planar mirror facet into a direction antiparallel to the $z$-axis.}\label{fig2}
	\end{figure}

\begin{figure}[ht!]\centering
		\includegraphics[width=0.45\textwidth]{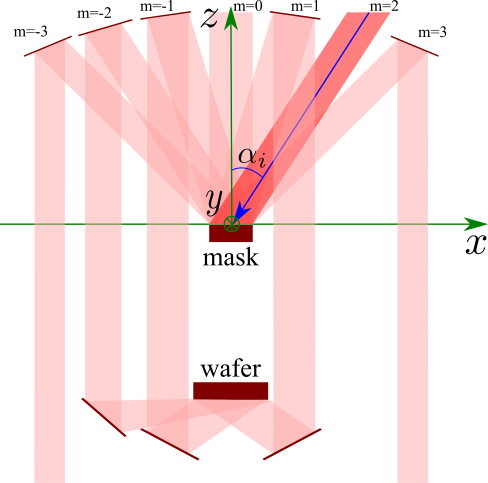}
		\caption{Schematic of the all-reflective two-mirror faceted projection architecture.}\label{fig3}
	\end{figure}

	Hence, the design task reduces to transforming the transverse spatial frequency of each $m$-th harmonic scattered by the mask into harmonic $n=-m$ on the wafer in the common Cartesian coordinate system:
	\begin{equation}
		m\kappa_x \longrightarrow -m\kappa_x^{(\mathrm{w})}.
	\end{equation}
	Because this spatial-frequency mapping is linear, it can be realized using a dual set of planar mirror facets. Specifically, the first mirror system redirects each $m$-th beam along a direction parallel to the $z$-axis (Fig.~\ref{fig2}), whereupon the second mirror system redirects it toward the wafer at an angle corresponding to $k_x = -m\kappa_x^{(\mathrm{w})}$ (Fig.~\ref{fig3}). Figure~\ref{fig3} schematically illustrates the all-reflective system comprising two sets of planar mirror facets. The first set converts the divergent beams scattered by the mask into a parallel array of vertical beams. The second set redirects a subset of these parallel beams at the required convergence angles onto the wafer.
	
	We note that because the illumination beam with spatial harmonic $\tilde{m}$ enters through an optical input aperture occupying the position corresponding to the $-\tilde{m}$ reflection order (back reflection), order $m = -\tilde{m}$ is omitted from the projected spectrum (Fig.~\ref{fig3}). Thus, at the first reflection, the amplitude $\tilde{A}_{-\tilde{m}}^{(r)}(x, z)$ is multiplied by zero.
	
	Referring to Fig.~\ref{fig2}, the beam parameters and mirror facet orientation for vertical reflection satisfy:
	\begin{align}
		& \cos \alpha_m = \frac{m \kappa_x}{k_0}, \notag\\
		& a_m = L \sin \alpha_m, \notag\\
		& \gamma_m = \frac{\pi}{4} - \frac{\alpha_m}{2},
	\end{align}
	where $\alpha_m$ is the angle of the $m$-th beam measured from the mask plane (horizontal), $a_m$ is the beam cross-sectional width, and $\gamma_m$ is the tilt angle of the facet ($\tau_m=\gamma_m$ in Fig.~\ref{fig2}). Each $m$-th beam projects an illuminated footprint $L_m^{(\mathrm{w})}$ on the wafer plane given by:
	\begin{equation}
		a_m = L_m^{(\mathrm{w})} \sin\alpha_m^{(\mathrm{w})},
	\end{equation}
	where $\sin\alpha_m^{(\mathrm{w})} = k_{z;m}^{(\mathrm{w})}/k_0$ and $\alpha_m^{(\mathrm{w})}$ is the grazing angle at the wafer plane. Consequently:
	\begin{equation}
		L_m^{(\mathrm{w})} = L \frac{k_{z;m}}{k_{z;m}^{(\mathrm{w})}} = L \left(\frac{1 - \left(\dfrac{m \kappa_x}{k_0}\right)^2}{1 - 16\left(\dfrac{m \kappa_x}{k_0}\right)^2}\right)^{1/2}.
	\end{equation}
	
	To prevent beam crosstalk, each facet of the first mirror must intercept only its assigned diffraction order. This requires that adjacent beam envelopes do not overlap at the mirror plane. The intersection point of the boundaries of adjacent orders $m$ and $m+1$ is derived from their ray equations:
	\begin{align}
		& z = \frac{k_{z;m}}{\kappa_x m} \left(x - \frac{L}{2}\right),\notag\\
		& z = \frac{k_{z;m+1}}{\kappa_x (m + 1)} \left(x + \frac{L}{2}\right),
	\end{align}
	for beams of diffraction orders $m$ and $m+1$, respectively. Equating the $z$-coordinates yields the intersection coordinates:
	\begin{align}
		x &= \frac{L}{2} \frac{(m+1)k_{z;m} + m k_{z;m+1}}{(m+1) k_{z;m} - m k_{z;m+1}},\label{eq9}\\
		z &= \frac{L}{\kappa_x} \frac{k_{z;m} k_{z;m+1}}{(m+1) k_{z;m} - m k_{z;m+1}}.\label{eq10}
	\end{align}
	
\begin{figure}[ht!]\centering
		\includegraphics[width=0.5\textwidth]{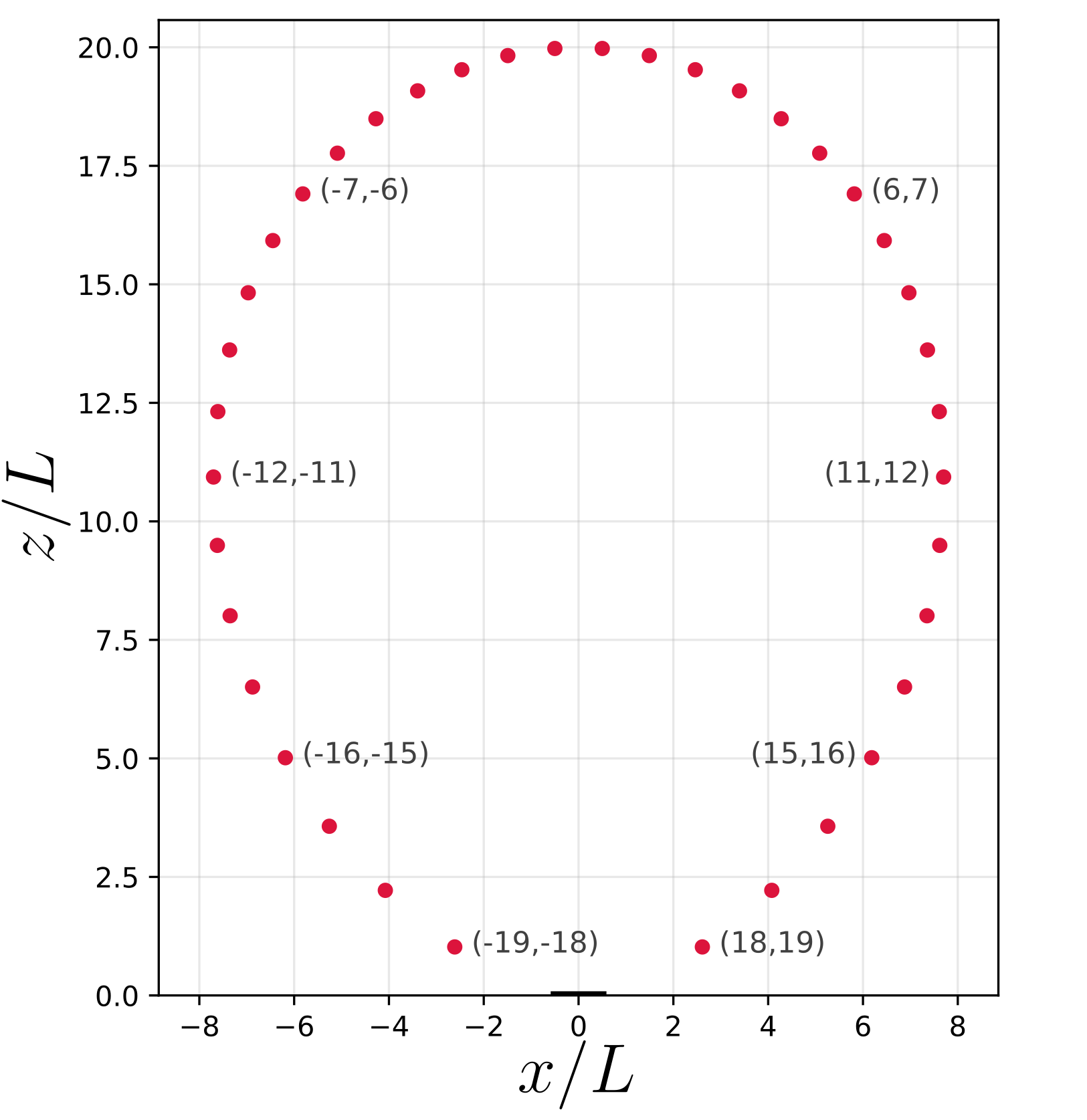}
		\caption{Spatial intersection coordinates of adjacent beam boundaries (markers) computed via Eqs.~\eqref{eq9} and~\eqref{eq10} for $L_x = 20\lambda$ in the $xOz$ plane. Thin solid lines denote beam edges originating from the mask boundaries ($|x| = L/2$, $z = 0$), the dashed line traces the intersection loci, and labels indicate diffraction order pairs $(m, m+1)$. Coordinates are normalized to mask width $L$.}\label{fig5}
	\end{figure}

	Figure~\ref{fig5} plots these boundary intersection coordinates calculated from Eqs.~\eqref{eq9} and~\eqref{eq10} for $L_x = 20\lambda$. Coordinates are normalized to $L$. Adjacent beams overlap below these intersection points and fully decouple above them. Hence, the first-mirror facets must be positioned at $z$-coordinates exceeding these critical thresholds.
	
\begin{figure}[ht!]\centering
		\includegraphics[width=0.45\textwidth]{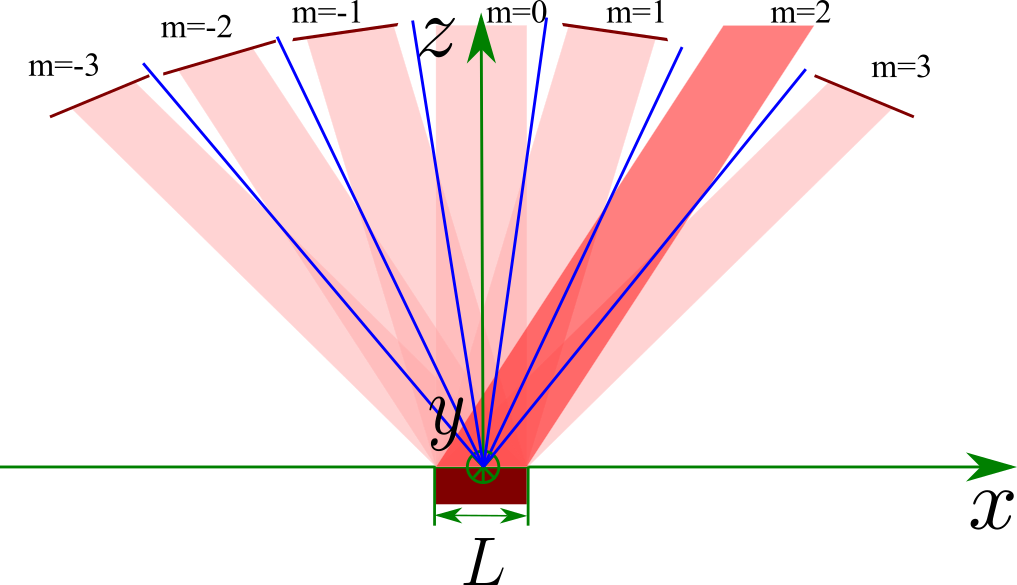}
		\caption{First-stage mirror system. Blue lines designate angular spatial-frequency sectors allocated to individual diffraction harmonics.}\label{fig4}
	\end{figure}

	The angular sector available for positioning the facet for order $m$ is bounded by rays connecting the mask center to the nearest adjacent intersection points, shown as blue boundary lines in Fig.~\ref{fig4}.
	
	\section{2D Two-Mirror Projection System}\label{2dmirrors}
	
	\subsection{Design of the Two-Mirror Projection System}
	
	The development of the two-mirror projection system proceeds through an iterative design progression:
	\begin{center}
		\begin{tikzpicture}[outer sep=auto]
			\draw[line width=5pt]
			(-1.2,0)  node (f0) {}
			(0,0)  node (f) {Suggestions}
			(3.8,0)  node (d) {Design of the mirror system}
			(8.8,0)  node (s) {Discussion of disadvantages}
			(10.8,0)  node (s0) {};
			
			\draw[->] (f) -- (d);
			\draw[->] (d) -- (s);
			\draw[overlay, arrows = {->[]}] ++(s0.east) -- +(0.2,0) |- +(0,-0.5) -| (f0.west) -- +(0.3,0);  
		\end{tikzpicture}
	\end{center}
	This approach enables systematic iterative refinement of the two-mirror projection system.
	
	Consider an illumination beam incident at $6^{\circ}$ relative to the $z$-axis (the industrial standard for EUV lithography). The mask lateral dimension is $L = 10$~mm, the exposure wavelength is $\lambda = 11.2$~nm, and the mask pattern period is $L_x = 8\lambda/\sin 6^{\circ} \approx 857$~nm. Under these conditions, the maximum propagating mask-order index is $M = \lfloor k_0/\kappa_x \rfloor = 76$, and the maximum wafer-order index under $4\times$ demagnification is $N = \lfloor M/4 \rfloor = 19$.
	
	In the configurations evaluated below, the first mirror is located at $z = Z_1 = 1000$~mm. The center of the $m$-th facet on the first mirror is located at $(\pm x_1^{(m)}, Z_1)$, where $x_1^{(m)} = Z_1 \tan\theta_m$ and $\tan \theta_m = \kappa_x m / k_{z;m}$.

	\subsubsection{Example 1: Mask and Wafer Located on the Same Side of the Second Mirror}
	
	In this baseline geometry, the second mirror system is positioned below the wafer plane (mask and wafer are on the same side of it). The tilt angle of the $m$-th facet on the first mirror relative to the horizontal in this and the following example is $\alpha_1^{(m)} = \theta_m / 2$. The center of the $m$-th facet on the second mirror is positioned at $(\pm x_1^{(m)}, z_f^{(m)})$, where $z_f^{(m)} = Z_{\mathrm{w}} - x_1^{(m)} / \tan\theta_m^{(\mathrm{w})}$ with $\tan \theta_m^{(\mathrm{w})} = 4 \kappa_x m / k_{z;m}^{(\mathrm{w})}$. The tilt angle of the second-mirror facet relative to the horizontal is $\alpha_2^{(m)} = \theta_m^{(\mathrm{w})} / 2$.
	
	Table~\ref{tab:mirrors} lists the structural parameters for the 19 positive-index transmitted orders.
	
	\begin{longtable}{ccccccccc}
		\caption{Mirror system configuration for Example 1 ($\lambda=11.2$~nm, $L=10$~mm, $L_x=857$~nm, $M=76$, $N=19$, $Z_1=1000$~mm, $Z_{\mathrm{w}}=-300$~mm).}\label{tab:mirrors}\\
		\toprule
		$m$ & $\theta_m$ (${}^\circ$) & $\theta_m^{(\mathrm{w})}$ (${}^\circ$) & $x_1^{(m)}$~(mm) & $z_f^{(m)}$~(mm) & $a_m$~(mm) & $L_m^{(\mathrm{w})}$~(mm) & $\alpha_1^{(m)}$ (${}^\circ$) & $\alpha_2^{(m)}$ (${}^\circ$)\\
		\midrule
		\endfirsthead
		\multicolumn{9}{c}{\tablename~\thetable~(Continued)}\\
		\toprule
		$m$ & $\theta_m$ (${}^\circ$) & $\theta_m^{(\mathrm{w})}$ (${}^\circ$) & $x_1^{(m)}$~(mm) & $z_f^{(m)}$~(mm) & $a_m$~(mm) & $L_m^{(\mathrm{w})}$~(mm) & $\alpha_1^{(m)}$ (${}^\circ$) & $\alpha_2^{(m)}$ (${}^\circ$)\\
		\midrule
		\endhead
		\bottomrule
		\endfoot
		\bottomrule
		\endlastfoot
		1 &  0.75 &  3.00 &   13.1 & $-$549.7 & 10.00 & 10.01 & 0.37 &  1.50\\
		2 &  1.50 &  6.00 &   26.1 & $-$548.7 & 10.00 & 10.05 & 0.75 &  3.00\\
		3 &  2.25 &  9.02 &   39.2 & $-$547.1 &  9.99 & 10.12 & 1.12 &  4.51\\
		4 &  3.00 & 12.07 &   52.3 & $-$544.8 &  9.99 & 10.21 & 1.50 &  6.03\\
		5 &  3.75 & 15.15 &   65.5 & $-$541.8 &  9.98 & 10.34 & 1.87 &  7.57\\
		6 &  4.50 & 18.28 &   78.6 & $-$538.1 &  9.97 & 10.50 & 2.25 &  9.14\\
		7 &  5.25 & 21.46 &   91.8 & $-$533.6 &  9.96 & 10.70 & 2.62 & 10.73\\
		8 &  6.00 & 24.72 &  105.1 & $-$528.3 &  9.95 & 10.95 & 3.00 & 12.36\\
		9 &  6.75 & 28.06 &  118.4 & $-$522.2 &  9.93 & 11.25 & 3.38 & 14.03\\
		10 &  7.51 & 31.51 &  131.8 & $-$515.0 &  9.91 & 11.63 & 3.75 & 15.75\\
		11 &  8.26 & 35.09 &  145.2 & $-$506.7 &  9.90 & 12.09 & 4.13 & 17.55\\
		12 &  9.02 & 38.84 &  158.8 & $-$497.2 &  9.88 & 12.68 & 4.51 & 19.42\\
		13 &  9.78 & 42.80 &  172.4 & $-$486.1 &  9.85 & 13.43 & 4.89 & 21.40\\
		14 & 10.54 & 47.03 &  186.1 & $-$473.3 &  9.83 & 14.42 & 5.27 & 23.51\\
		15 & 11.30 & 51.62 &  199.9 & $-$458.3 &  9.81 & 15.80 & 5.65 & 25.81\\
		16 & 12.07 & 56.74 &  213.8 & $-$440.2 &  9.78 & 17.83 & 6.03 & 28.37\\
		17 & 12.83 & 62.68 &  227.8 & $-$417.7 &  9.75 & 21.25 & 6.42 & 31.34\\
		18 & 13.60 & 70.18 &  242.0 & $-$387.2 &  9.72 & 28.66 & 6.80 & 35.09\\
		19 & 14.37 & 83.23 &  256.3 & $-$330.4 &  9.69 & 82.13 & 7.19 & 41.61\\
	\end{longtable}
	
	The lateral span of the first mirror along the $x$-axis is $\pm 256.3$~mm at fixed height $z = Z_1 = 1000$~mm, i.e., all facet centers of the first mirror lie in a single horizontal plane (facet tilt angles range from $0.37^\circ$ to $7.19^\circ$, visually indistinguishable from a flat mirror). The second mirror spans an identical lateral range, but its facets lie along a curved profile $z_f^{(m)} = Z_{\mathrm{w}} - x_1^{(m)} / \tan\theta_m^{(\mathrm{w})}$ with heights spanning $z_f^{(1)} = -549.7$~mm (center) to $z_f^{(19)} = -330.4$~mm (edge). The maximum numerical aperture of the system reaches $\mathrm{NA}_{\max} = \sin\theta_{19}^{(\mathrm{w})} = 0.993$.
	
	\begin{figure}[ht!]\centering
		\includegraphics[width=0.3\textwidth,keepaspectratio]{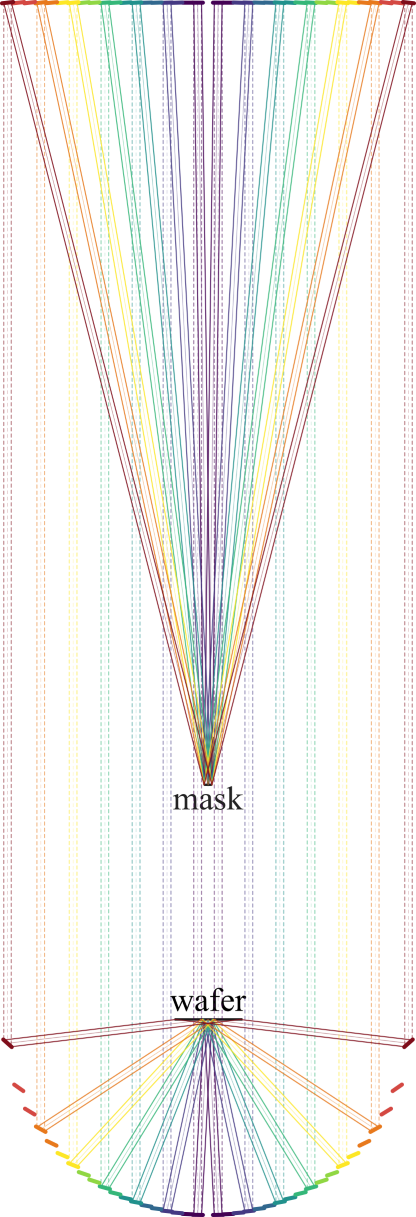}
		\caption{Ray tracing in the all-reflective two-mirror projection system for $N=19$ orders (showing orders $m = 1, 4, 7, 10, 13, 16, 19$). Color represents diffraction order: purple denotes $m=1$ ($\theta_m^{(\mathrm{w})} = 3.0^\circ$), dark red denotes $m=19$ ($\theta_m^{(\mathrm{w})} = 83.2^\circ$). Solid lines show marginal rays from the mask edges $\pm L/2$. Dashed lines show axial rays. Rays trace from the mask ($z=0$) through mirror 1 ($z=1000$~mm) and mirror 2 ($z \in [-550, -330]$~mm) to the wafer ($z=-300$~mm).}\label{fig:rays}
	\end{figure}
	
	Figure~\ref{fig:rays} displays the ray-trace diagram in the developed projection system. Each $m$-th harmonic of the field scattered by the mask is converted by the first set of $2N$ planar facets into a beam parallel to the $z$-axis, and then focused onto the wafer at angle $\theta_m^{(\mathrm{w})}$ to the normal by the second set of $2N$ facets. The beam width $a_m \approx 10$~mm is preserved upon first reflection (i.e., redirection into the $-z$ direction), but after reflection from the second (tilted) mirror it projects onto the wafer with a footprint enlarged by a factor of $1/\cos\theta_m^{(\mathrm{w})}$. For $m=19$ ($\theta_{19}^{(\mathrm{w})} = 83.2^\circ$), the footprint reaches $82.1$~mm. As the beam arrives at near-grazing incidence on the wafer, the perpendicular cross-section $a_m = 9.7$~mm is stretched by a factor of 8.5.
	
	\textbf{Discussion.} The analyzed system forms a demagnified pattern on the wafer via spatial harmonic transformation. The primary limitation of this design is that the wafer is situated within the ray convergence zone directed by the second mirror, severely restricting the physical clearance required for wafer stage manipulation and handling --- a critical constraint in industrial lithography. Furthermore, the $m=0$ beam is omitted from the field arriving at the wafer, and high-order beams produce relatively large footprints on the wafer plane. The latter two aspects are secondary compared to the first constraint.
	
	\textbf{Suggestion 1.} Position the wafer outside the space between the first and second mirrors.

	\subsubsection{Example 2: Wafer Positioned Below the Second Mirror}
	
	We now consider a configuration in which the second mirror system is positioned between the mask and the wafer, placing the wafer entirely below the second mirror (mask and wafer are on opposite sides of it). The first mirror remains identical ($z = Z_1 = 1000$~mm). The vertically descending beams are reflected by the second mirror toward the center of the wafer while continuing downward. The center of the $m$-th facet of the second mirror is positioned at $(\pm x_1^{(m)}, z_f^{(m)})$, where
	\begin{equation}
		z_f^{(m)} = Z_{\mathrm{w}} + \frac{x_1^{(m)}}{\tan\theta_m^{(\mathrm{w})}},
	\end{equation}
	so that the second mirror system spans the domain from $z_f^{(1)} = -50.3$~mm (center) to $z_f^{(19)} = -269.6$~mm (edge). The height difference ($219$~mm) is identical to that in Example 1, but the second mirror system is now positioned above the wafer. The facet tilt angle relative to the horizontal is $\alpha_2^{(m)} = 90^\circ - \theta_m^{(\mathrm{w})} / 2$: the facets are nearly vertical, and beams strike them at grazing angles $\theta_m^{(\mathrm{w})}/2$ measured from the facet plane (from $1.5^\circ$ for $m=1$ to $41.6^\circ$ for $m=19$), corresponding to incidence angles $90^\circ - \theta_m^{(\mathrm{w})}/2$ from the facet normal.
	
\begin{figure}[ht!]\centering
		\includegraphics[width=0.3\textwidth,keepaspectratio]{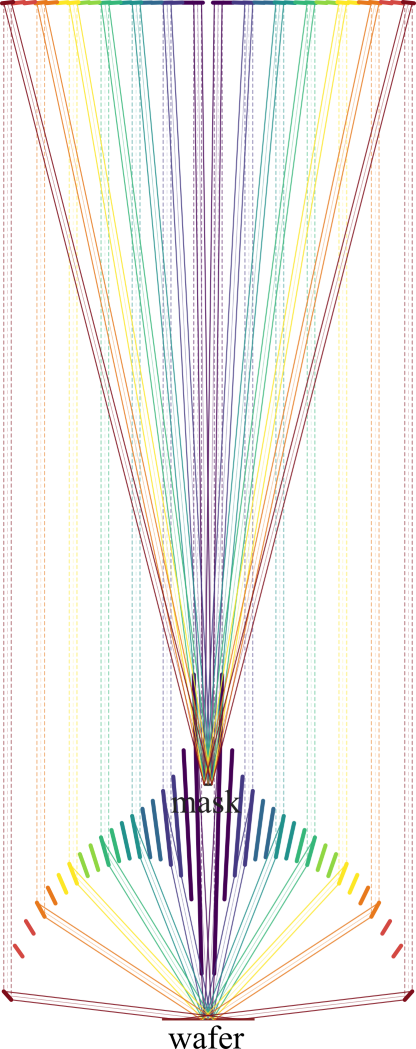}
		\caption{Ray tracing in the configuration with the wafer below the second mirror ($N=19$, showing orders $m = 1, 4, 7, 10, 13, 16, 19$). The second mirror system is positioned between the mask ($z=0$) and wafer ($z=-300$~mm). Its facets are nearly vertical (tilt relative to the horizontal $90^\circ - \theta_m^{(\mathrm{w})}/2$) and plotted with full length $\ell_2^{(m)} = a_m/\sin(\theta_m^{(\mathrm{w})}/2)$ required to intercept the entire beam (up to $382.5$~mm for $m=1$).}\label{fig:rays2}
	\end{figure}

	Table~\ref{tab:mirrors2} details the mirror configurations for the 19 positive-index orders. The final column $\ell_2^{(m)}$ defines the minimum facet length required to intercept the full beam width $a_m$ at grazing incidence: $\ell_2^{(m)} = a_m / \sin(\theta_m^{(\mathrm{w})}/2)$. The ray trace is shown in Fig.~\ref{fig:rays2}.
	
	\begingroup\setlength{\tabcolsep}{3pt}
\begin{longtable}{cccccccccc}
		\caption{Mirror system parameters for the configuration with the wafer below the second mirror ($\lambda=11.2$~nm, $L=10$~mm, $L_x=857$~nm, $M=76$, $N=19$, $Z_1=1000$~mm, $Z_{\mathrm{w}}=-300$~mm).}\label{tab:mirrors2}\\
		\toprule
		$m$ & $\theta_m$ (${}^\circ$) & $\theta_m^{(\mathrm{w})}$ (${}^\circ$) & $x_1^{(m)}$~(mm) & $z_f^{(m)}$~(mm) & $a_m$~(mm) & $L_m^{(\mathrm{w})}$~(mm) & $\alpha_1^{(m)}$ (${}^\circ$) & $\alpha_2^{(m)}$ (${}^\circ$) & $\ell_2^{(m)}$~(mm)\\
		\midrule
		\endfirsthead
		\multicolumn{10}{c}{\tablename~\thetable~(Continued)}\\
		\toprule
		$m$ & $\theta_m$ (${}^\circ$) & $\theta_m^{(\mathrm{w})}$ (${}^\circ$) & $x_1^{(m)}$~(mm) & $z_f^{(m)}$~(mm) & $a_m$~(mm) & $L_m^{(\mathrm{w})}$~(mm) & $\alpha_1^{(m)}$ (${}^\circ$) & $\alpha_2^{(m)}$ (${}^\circ$) & $\ell_2^{(m)}$~(mm)\\
		\midrule
		\endhead
		\bottomrule
		\endfoot
		\bottomrule
		\endlastfoot
		1 &  0.75 &  3.00 &   13.1 & $-$50.3 & 10.00 & 10.01 & 0.37 & 88.50 & 382.5\\
		2 &  1.50 &  6.00 &   26.1 & $-$51.3 & 10.00 & 10.05 & 0.75 & 87.00 & 191.0\\
		3 &  2.25 &  9.02 &   39.2 & $-$52.9 &  9.99 & 10.12 & 1.12 & 85.49 & 127.1\\
		4 &  3.00 & 12.07 &   52.3 & $-$55.2 &  9.99 & 10.21 & 1.50 & 83.97 &  95.0\\
		5 &  3.75 & 15.15 &   65.5 & $-$58.2 &  9.98 & 10.34 & 1.87 & 82.43 &  75.7\\
		6 &  4.50 & 18.28 &   78.6 & $-$61.9 &  9.97 & 10.50 & 2.25 & 80.86 &  62.8\\
		7 &  5.25 & 21.46 &   91.8 & $-$66.4 &  9.96 & 10.70 & 2.62 & 79.27 &  53.5\\
		8 &  6.00 & 24.72 &  105.1 & $-$71.7 &  9.95 & 10.95 & 3.00 & 77.64 &  46.5\\
		9 &  6.75 & 28.06 &  118.4 & $-$77.8 &  9.93 & 11.25 & 3.38 & 75.97 &  41.0\\
		10 &  7.51 & 31.51 &  131.8 & $-$85.0 &  9.91 & 11.63 & 3.75 & 74.25 &  36.5\\
		11 &  8.26 & 35.09 &  145.2 & $-$93.3 &  9.90 & 12.09 & 4.13 & 72.45 &  32.8\\
		12 &  9.02 & 38.84 &  158.8 & $-$102.8 &  9.88 & 12.68 & 4.51 & 70.58 &  29.7\\
		13 &  9.78 & 42.80 &  172.4 & $-$113.9 &  9.85 & 13.43 & 4.89 & 68.60 &  27.0\\
		14 & 10.54 & 47.03 &  186.1 & $-$126.7 &  9.83 & 14.42 & 5.27 & 66.49 &  24.6\\
		15 & 11.30 & 51.62 &  199.9 & $-$141.7 &  9.81 & 15.80 & 5.65 & 64.19 &  22.5\\
		16 & 12.07 & 56.74 &  213.8 & $-$159.8 &  9.78 & 17.83 & 6.03 & 61.63 &  20.6\\
		17 & 12.83 & 62.68 &  227.8 & $-$182.3 &  9.75 & 21.25 & 6.42 & 58.66 &  18.7\\
		18 & 13.60 & 70.18 &  242.0 & $-$212.8 &  9.72 & 28.66 & 6.80 & 54.91 &  16.9\\
		19 & 14.37 & 83.23 &  256.3 & $-$269.6 &  9.69 & 82.13 & 7.19 & 48.39 &  14.6\\
	\end{longtable}
\endgroup
	
	\textbf{Discussion.} The lateral span of the mirrors along the $x$-axis ($\pm 256.3$~mm) and the maximum numerical aperture $\mathrm{NA}_{\max} = 0.993$ remain identical to Example 1. The beam widths $a_m$ and footprint dimensions $L_m^{(\mathrm{w})}$ on the wafer are unchanged. The key difference of this configuration is grazing incidence on the second mirror. At sufficiently small grazing angles, a suitable material can provide useful total-external-reflection reflectance without a Bragg stack. The critical angle and absorption depend on the material and wavelength and must be evaluated from the complex refractive index. No universal $15^\circ$ threshold is assumed. At larger grazing angles, multilayer coatings require optimization for the actual incidence angle. The total optical path length of the central ray from the mask to the wafer increases from $2300.4$~mm ($m=1$) to $2560.0$~mm ($m=19$), corresponding to an optical path difference of $\approx 260$~mm.
	
	The minimum facet length $\ell_2^{(m)}=a_m/\sin(\theta_m^{(\mathrm{w})}/2)$ is large for low diffraction orders, reaching $382.5$~mm for $m=1$, $191.0$~mm for $m=2$, and $127.1$~mm for $m=3$. This follows from the grazing-incidence footprint enlargement by $1/\sin(\theta_m^{(\mathrm{w})}/2)$.
	
	\textbf{Suggestion 2.} To obtain more compact second-mirror facets, place the wafer in the base plane of the first mirror and the mask in the base plane of the second. Furthermore, this layout restores the $m=0$ beam to the field reaching the wafer.

	\subsubsection{Example 3: Wafer in the Plane of Mirror 1, Mask in the Base Plane of Mirror 2}
	
\begin{figure}[ht!]\centering
		\includegraphics[width=0.5\textwidth,keepaspectratio]{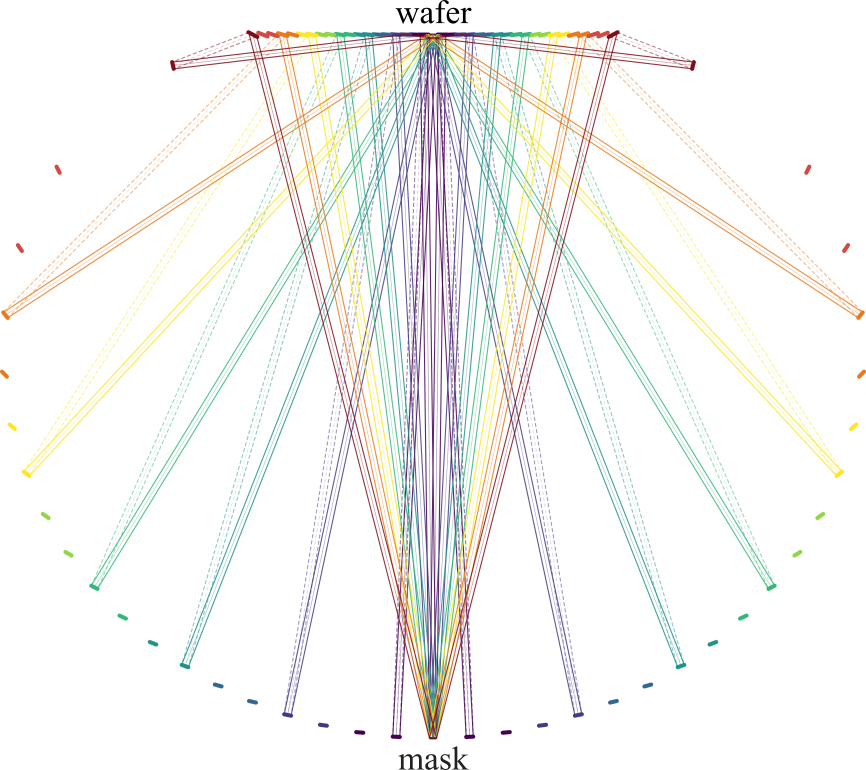}
		\caption{Ray tracing with the wafer in the plane of the first mirror and the mask at the base of the second ($N=19$, showing orders $m = 1, 4, 7, 10, 13, 16, 19$). The beam from the mask ($z=0$) propagates upward to mirror 1 ($z=1000$~mm), reflects downward and outward at angle $(\theta_m^{(\mathrm{w})} - \theta_m)$, intercepts the second mirror facet ($z_f$ from $2$~mm to $956$~mm), which directs the beam upward to the center of the wafer ($z = Z_1$) at convergence angle $\theta_m^{(\mathrm{w})}$.}\label{fig:rays3}
	\end{figure}

	In this example, the wafer is placed in the plane of the first mirror ($z = Z_1 = 1000$~mm) and the mask is located at the base of the second ($z = 0$). The second mirror consists of a system of facets distributed in height: central facets (small $m$) reside near the mask, while peripheral facets (large $m$) ascend toward the wafer. The spatial distance from the wafer center to the central facets of the second mirror is on the order of the mask--wafer distance $Z_1 = 1000$~mm. Figure~\ref{fig:rays3} shows the ray trace for this projection system.
	
	First mirror at $z = Z_1$: facet $m$ is centered at $(\pm x_1^{(m)}, Z_1)$, $x_1^{(m)} = Z_1\tan\theta_m$. Unlike Example 1, the first mirror does not collimate the beams vertically, but deflects them downward and outward:
	\begin{equation}
		(\sin\theta_m, \cos\theta_m) \longrightarrow \left[\sin(\theta_m^{(\mathrm{w})} - \theta_m),\, -\cos(\theta_m^{(\mathrm{w})} - \theta_m)\right].
	\end{equation}
	The angle of incidence on the first mirror is $i_1^{(m)} = \theta_m^{(\mathrm{w})}/2$ ($1.5^{\circ}$--$41.6^{\circ}$), and the facet tilt is $\alpha_1^{(m)} = \theta_m^{(\mathrm{w})}/2 - \theta_m$ ($0.8^{\circ}$--$27.2^{\circ}$ relative to the horizontal).
	
	Second mirror: facet $m$ is centered at $(\pm x_f^{(m)}, z_f^{(m)})$, where
	\begin{equation}
		z_f^{(m)} = Z_1 - \frac{Z_1 \tan\theta_m}{\tan\theta_m^{(\mathrm{w})} - \tan(\theta_m^{(\mathrm{w})} - \theta_m)}, \qquad
		x_f^{(m)} = (Z_1 - z_f^{(m)})\tan\theta_m^{(\mathrm{w})}.
	\end{equation}
	For small $m$, $\tan\theta_m \ll \tan\theta_m^{(\mathrm{w})}$, yielding $z_f^{(m)} \approx 0$ --- central facets lie near the mask. For $m = 19$, $z_f \approx 956$~mm --- the facet is near the wafer. The second mirror redirects the beam toward the center of the wafer:
	\begin{equation}
		\left[\pm\sin(\theta_m^{(\mathrm{w})} - \theta_m), -\cos(\theta_m^{(\mathrm{w})} - \theta_m)\right] \longrightarrow \left[\mp\sin\theta_m^{(\mathrm{w})}, \cos\theta_m^{(\mathrm{w})}\right].
	\end{equation}
	The angle of incidence on the second mirror is $i_2^{(m)} = \theta_m/2$ ($0.4^\circ$--$7.2^\circ$), and facet tilt angles are $\alpha_2^{(m)}$ ($2.6^\circ$--$76.0^\circ$ relative to the horizontal). The facet lengths of the second mirror are compact: $\ell_2^{(m)} = a_m/\cos(\theta_m/2) \approx 10$~mm. Geometric configurations for all facets $m = 1$--$19$ are listed in Table~\ref{tab:mirrors7}.
	
	{\small
		\begin{longtable}{cccccccccc}
			\caption{Mirror system configuration with the wafer in the plane of the first mirror ($i_1$ and $i_2$ are angles of incidence on the first and second mirrors, respectively, $P_m$ is total optical path length).}\label{tab:mirrors7}\\
			\toprule
			$m$ & $\theta_m$ (${}^\circ$) & $\theta_m^{(\mathrm{w})}$ (${}^\circ$) & $x_1^{(m)}$~(mm) & $x_f^{(m)}$~(mm) & $z_f^{(m)}$~(mm) & $i_1^{(m)}$ (${}^\circ$) & $i_2^{(m)}$ (${}^\circ$) & $\ell_2^{(m)}$~(mm) & $P_m$~(mm)\\
			\midrule
			\endfirsthead
			\multicolumn{10}{c}{\tablename~\thetable~(Continued)}\\
			\toprule
			$m$ & $\theta_m$ (${}^\circ$) & $\theta_m^{(\mathrm{w})}$ (${}^\circ$) & $x_1^{(m)}$~(mm) & $x_f^{(m)}$~(mm) & $z_f^{(m)}$~(mm) & $i_1^{(m)}$ (${}^\circ$) & $i_2^{(m)}$ (${}^\circ$) & $\ell_2^{(m)}$~(mm) & $P_m$~(mm)\\
			\midrule
			\endhead
			\bottomrule
			\endfoot
			\bottomrule
			\endlastfoot
			1 &  0.75 &  3.00 &   13.1 &   52.2 &    2.0 &  1.50 &  0.37 &   10.0 & 2998.1\\
			2 &  1.50 &  6.00 &   26.1 &  104.2 &    8.2 &  3.00 &  0.75 &   10.0 & 2992.5\\
			3 &  2.25 &  9.02 &   39.2 &  155.8 &   18.5 &  4.51 &  1.12 &   10.0 & 2982.9\\
			4 &  3.00 & 12.07 &   52.3 &  206.7 &   33.0 &  6.03 &  1.50 &   10.0 & 2969.5\\
			5 &  3.75 & 15.15 &   65.5 &  256.7 &   51.8 &  7.57 &  1.87 &   10.0 & 2951.8\\
			6 &  4.50 & 18.28 &   78.6 &  305.5 &   74.9 &  9.14 &  2.25 &   10.0 & 2929.8\\
			7 &  5.25 & 21.46 &   91.8 &  352.8 &  102.6 & 10.73 &  2.62 &   10.0 & 2903.1\\
			8 &  6.00 & 24.72 &  105.1 &  398.2 &  134.9 & 12.36 &  3.00 &   10.0 & 2871.2\\
			9 &  6.75 & 28.06 &  118.4 &  441.3 &  172.1 & 14.03 &  3.38 &    9.9 & 2833.8\\
			10 &  7.51 & 31.51 &  131.8 &  481.6 &  214.4 & 15.75 &  3.75 &    9.9 & 2790.0\\
			11 &  8.26 & 35.09 &  145.2 &  518.4 &  262.2 & 17.55 &  4.13 &    9.9 & 2739.0\\
			12 &  9.02 & 38.84 &  158.8 &  550.9 &  315.8 & 19.42 &  4.51 &    9.9 & 2679.6\\
			13 &  9.78 & 42.80 &  172.4 &  578.1 &  375.7 & 21.40 &  4.89 &    9.9 & 2610.1\\
			14 & 10.54 & 47.03 &  186.1 &  598.4 &  442.6 & 23.51 &  5.27 &    9.9 & 2528.3\\
			15 & 11.30 & 51.62 &  199.9 &  609.5 &  517.3 & 25.81 &  5.65 &    9.9 & 2430.4\\
			16 & 12.07 & 56.74 &  213.8 &  608.1 &  601.2 & 28.37 &  6.03 &    9.8 & 2310.5\\
			17 & 12.83 & 62.68 &  227.8 &  587.6 &  696.5 & 31.34 &  6.42 &    9.8 & 2157.6\\
			18 & 13.60 & 70.18 &  242.0 &  533.1 &  807.8 & 35.09 &  6.80 &    9.8 & 1944.5\\
			19 & 14.37 & 83.23 &  256.3 &  369.8 &  956.1 & 41.61 &  7.19 &    9.8 & 1526.5\\
		\end{longtable}
	}
	
	\textbf{Discussion.} The optical path length decreases with order index: from $2998$~mm ($m=1$) to $1527$~mm ($m=19$), creating an optical path difference of $1472$~mm.
	
	\textbf{Suggestion 3.} Position the facets of the second mirror such that the optical path lengths across all diffraction orders are identical.

	\subsubsection{Example 4: Equal Optical Path Length, $\mathrm{NA} \sim 1$}
	
	In Example 3, optical path lengths $P_m$ vary substantially across orders (from $2998$~mm for $m=1$ to $1527$~mm for $m=19$), which introduces order-dependent propagation phase shifts even under coherent monochromatic illumination. In this example, the facet coordinates of the first mirror are preserved as in Example 3 ($z_1 = Z_1 = \mathrm{const}$, $x_1^{(m)} = Z_1\tan\theta_m$), while the facet heights $z_f^{(m)}$ of the second mirror and the first-mirror deflection angles $\beta_m$ are adjusted so that the total optical path length $P_m$ is identical for all diffraction orders (excluding $m=0$). The maximum positive order index is $N = 19$, and $\mathrm{NA}_{\max} = 0.993$, as in Example 3.
	
	The second mirror redirects each beam toward the center of the wafer $(0, Z_1)$ at angle $\theta_m^{(\mathrm{w})}$. The lateral coordinate is $x_f^{(m)} = (Z_1 - z_f^{(m)})\tan\theta_m^{(\mathrm{w})}$. The first mirror reflects the beam at a variable angle $\beta_m$ (from $2.2^\circ$ for $m=1$ to $81.2^\circ$ for $m=19$), determined by the position of the corresponding facet of the second mirror:
	\begin{equation}
		\tan\beta_m = \frac{x_f^{(m)} - x_1^{(m)}}{Z_1 - z_f^{(m)}}.
	\end{equation}
	The total optical path length is:
	\begin{equation}
		P_m = \frac{Z_1}{\cos\theta_m} + \left[\bigl(x_f^{(m)} - x_1^{(m)}\bigr)^2 + \bigl(Z_1 - z_f^{(m)}\bigr)^2\right]^{1/2} + \frac{Z_1 - z_f^{(m)}}{\cos\theta_m^{(\mathrm{w})}}.
	\end{equation}
	Here, the first term represents the path from the mask to the first mirror (fixed height $Z_1$, but varying $\cos\theta_m$ due to differing $k_{z;m}$). The second term is the path from the first to the second mirror. The third term is the path from the second mirror to the wafer. The parameter $u_m = Z_1 - z_f^{(m)}$ is determined via bisection for each order $m$ to enforce $P_m \approx 2998.1$~mm. As a result, $\Delta P = P_{\max} - P_{\min} = 0$ (within numerical precision), removing order-dependent propagation phase shifts. Phase differences introduced by the mask and coatings remain. For a source with finite bandwidth, temporal coherence requires a separate assessment.
	
	{\small
		\begin{longtable}{ccccccccc}
			\caption{Configuration of the equal-path mirror system ($N=19$, $\mathrm{NA} = 0.993$). Here $\beta$ is the reflection angle from mirror 1, $i_1$ and $i_2$ are angles of incidence, $\ell_2$ is the facet length of mirror 2, and $P_m$ is the total optical path length.}\label{tab:mirrors4}\\
			\toprule
			$m$ & $\theta_m^{(\mathrm{w})}$ (${}^\circ$) & $z_f$~(mm) & $x_f$~(mm) & $\beta$ (${}^\circ$) & $i_1$ (${}^\circ$) & $i_2$ (${}^\circ$) & $\ell_2$~(mm) & $P_m$~(mm)\\
			\midrule
			\endfirsthead
			\multicolumn{9}{c}{\tablename~\thetable~(Continued)}\\
			\toprule
			$m$ & $\theta_m^{(\mathrm{w})}$ (${}^\circ$) & $z_f$~(mm) & $x_f$~(mm) & $\beta$ (${}^\circ$) & $i_1$ (${}^\circ$) & $i_2$ (${}^\circ$) & $\ell_2$~(mm) & $P_m$~(mm)\\
			\midrule
			\endhead
			\bottomrule
			\endfoot
			\bottomrule
			\endlastfoot
			1 &  3.00 &    2.0 &   52.2 &  2.2 &  1.50 &  0.37 &  10.0 & 2998.1\\
			2 &  6.00 &    5.4 &  104.5 &  4.5 &  3.00 &  0.75 &  10.0 & 2998.1\\
			3 &  9.02 &   11.0 &  157.0 &  6.8 &  4.52 &  1.11 &  10.0 & 2998.1\\
			4 & 12.07 &   19.0 &  209.7 &  9.1 &  6.06 &  1.48 &  10.0 & 2998.1\\
			5 & 15.15 &   29.4 &  262.8 & 11.5 &  7.62 &  1.83 &  10.0 & 2998.1\\
			6 & 18.28 &   42.4 &  316.2 & 13.9 &  9.22 &  2.17 &  10.0 & 2998.1\\
			7 & 21.46 &   58.3 &  370.2 & 16.5 & 10.86 &  2.50 &  10.0 & 2998.1\\
			8 & 24.72 &   77.1 &  424.8 & 19.1 & 12.55 &  2.80 &  10.0 & 2998.1\\
			9 & 28.06 &   99.4 &  480.1 & 21.9 & 14.32 &  3.09 &   9.9 & 2998.1\\
			10 & 31.51 &  125.4 &  536.2 & 24.8 & 16.16 &  3.35 &   9.9 & 2998.1\\
			11 & 35.09 &  155.7 &  593.2 & 28.0 & 18.11 &  3.57 &   9.9 & 2998.1\\
			12 & 38.84 &  191.1 &  651.3 & 31.3 & 20.18 &  3.75 &   9.9 & 2998.1\\
			13 & 42.80 &  232.5 &  710.7 & 35.0 & 22.41 &  3.88 &   9.9 & 2998.1\\
			14 & 47.03 &  281.4 &  771.4 & 39.2 & 24.85 &  3.93 &   9.9 & 2998.1\\
			15 & 51.62 &  339.9 &  833.6 & 43.8 & 27.57 &  3.90 &   9.8 & 2998.1\\
			16 & 56.74 &  411.4 &  897.5 & 49.3 & 30.67 &  3.73 &   9.8 & 2998.1\\
			17 & 62.68 &  502.4 &  963.5 & 55.9 & 34.38 &  3.38 &   9.8 & 2998.1\\
			18 & 70.18 &  628.2 & 1031.6 & 64.8 & 39.19 &  2.70 &   9.7 & 2998.1\\
			19 & 83.23 &  869.1 & 1102.1 & 81.2 & 47.79 &  1.01 &   9.7 & 2998.1\\
		\end{longtable}
	}
	
\begin{figure}[ht!]\centering
		\includegraphics[width=\textwidth]{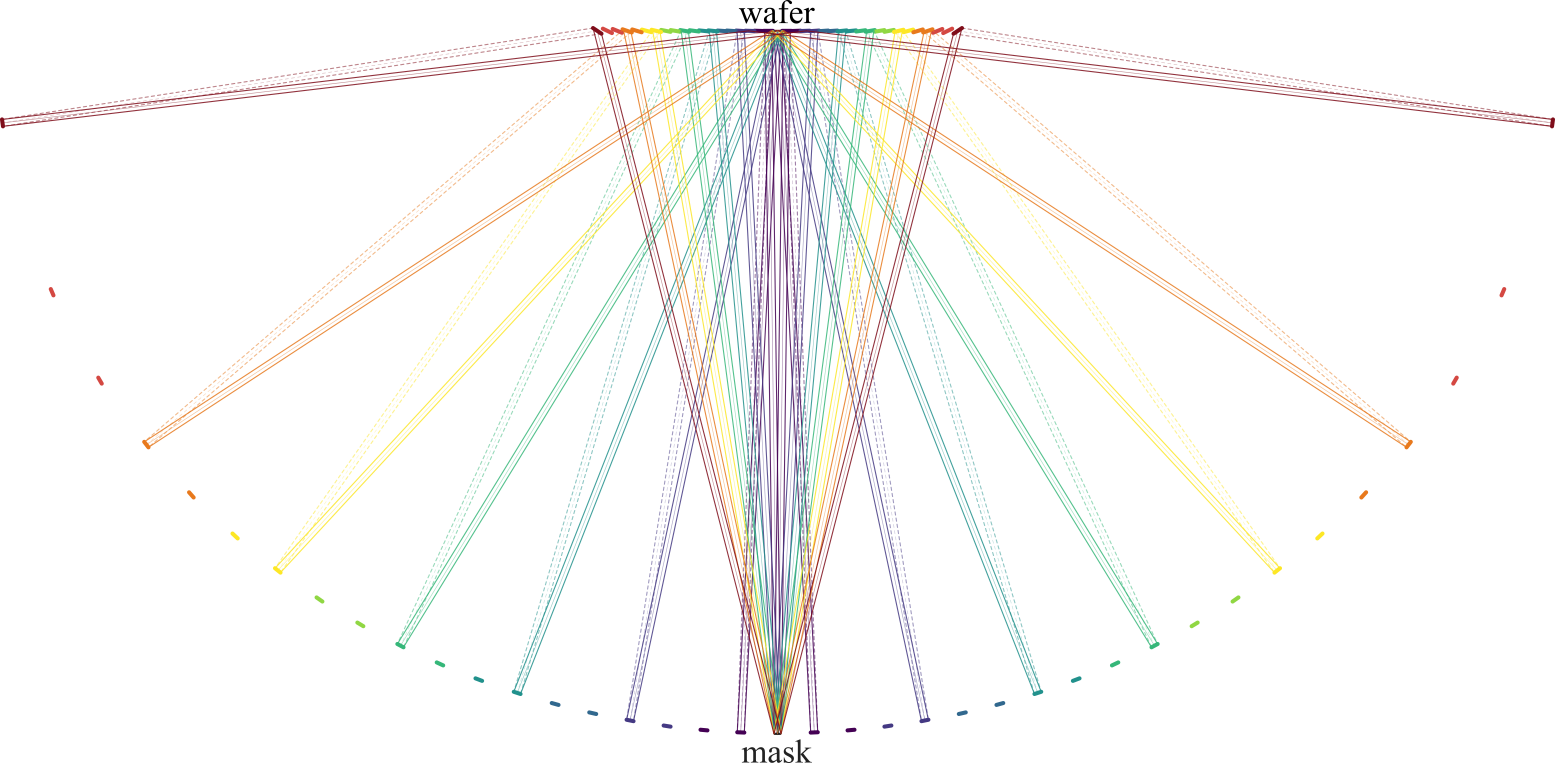}
		\caption{Ray tracing in the equal-path projection system ($N=19$, $\mathrm{NA} = 0.993$, showing orders $m = 1, 4, 7, 10, 13, 16, 19$). The first mirror lies in the plane $z = Z_1 = 1000$~mm. The deflection angle $\beta_m$ varies from $2.2^\circ$ to $81.2^\circ$. The second mirror is profiled such that total optical path length $P_m \approx 2998.1$~mm across all orders.}\label{fig:rays4}
	\end{figure}

	The angles of incidence on the first mirror, $i_1 = (\theta_m + \beta_m)/2$, vary from $1.5^\circ$ to $47.8^\circ$. On the second mirror, $i_2$ varies from $0.4^\circ$ to $3.9^\circ$, which is substantially smaller than $i_2 = \theta_m/2$ ($0.4^\circ$--$7.2^\circ$) in Example 3, since adjusting $\beta_m$ reduces the incidence angle on the second mirror for high orders. Facet lengths are $\ell_2 \approx 10$~mm. Figure~\ref{fig:rays4} shows the ray trace of the proposed equal-path projection system.
	
	\subsection{Bragg Coatings of Mirror Facets for Example 4}
	
	Each of the $2N = 38$ facets of the first and second mirrors in Example 4 operates at its own specific angle of incidence. Specifically, the angle of incidence on the first mirror $i_1^{(m)}$ ranges from $1.50^\circ$ ($m=1$) to $47.79^\circ$ ($m=19$), while $i_2^{(m)}$ on the second mirror ranges from $0.37^\circ$ to $3.93^\circ$ (see Tables~\ref{tab:bragg_13_5nm_MoSi} and~\ref{tab:bragg_11_2nm_RuBe}, column $\theta_{\mathrm{inc}}$). A standard Mo/Si Bragg coating (40--50 bilayers) provides maximum reflectance near a single fixed design angle. A single uniform coating cannot maintain high reflectance across the entire $0.4^\circ$--$47.8^\circ$ angular range. Therefore, individually optimized Bragg coatings maximizing reflectance at the exact operational angle of incidence must be designed for each facet.

	\subsubsection{Formulation of the Facet Bragg Optimization Problem} 
	For each facet specified by mirror index $j \in \{1,2\}$ and diffraction order $m \in \{1,\ldots,19\}$, a target angle of incidence $\theta_{\mathrm{target}} = i_j^{(m)}$ (measured from the facet normal) is assigned. We seek to determine the layer thicknesses $d_1$ (absorbing material) and $d_2$ (spacer material) of a periodic Bragg mirror comprising $N_b = 30$ bilayers (i.e., $2N_b = 60$ individual layers total) that maximize the reflectance $R = |r|^2$ at incident angle $\theta_{\mathrm{target}}$:
	\begin{equation}
		(d_1^*, d_2^*) = \arg\max_{d_1, d_2} \; R(d_1, d_2; \, \theta_{\mathrm{target}}).
		\label{eq:bragg_opt}
	\end{equation}
	Here, $r = r(d_1, d_2; \, \theta)$ is the complex field reflection coefficient calculated via the transfer-matrix method (TMM) for TE polarization. The multilayer structure is: vacuum / layer~1($d_1$) / layer~2($d_2$) / \ldots / layer~1($d_1$) / layer~2($d_2$) / vacuum (60 layers total, bounded by vacuum with $\eps = 1$ on both sides). Optical constants ($\eps = (1 - \delta - \mathrm{i}\beta)^2$, where $\delta$ is the refractive index decrement and $\beta$ is the absorption index) are obtained from tabulated reference data~\cite{HENKE1993181,CenterXRayOpt}.
	
	\textbf{Transfer-matrix method (TMM).} At incidence angle $\theta$, layer $i$ has longitudinal wavenumber $k_z^{(i)}=k_0(\eps_i-\sin^2\theta)^{1/2}$ and TE admittance $p_i=k_z^{(i)}/k_0$. The Fresnel coefficients are $r_{ij}=(p_i-p_j)/(p_i+p_j)$ and $t_{ij}=2p_i/(p_i+p_j)$. Define the entering-interface matrix and the propagation matrix of layer $i$ by
	\begin{equation}
	\begin{aligned}
	\mathbf D_{i-1,i}&=\frac{1}{t_{i-1,i}}\begin{pmatrix}1&r_{i-1,i}\\r_{i-1,i}&1\end{pmatrix},\\
	\mathbf P_i&=\operatorname{diag}(\mathrm e^{\mathrm i\phi_i},\mathrm e^{-\mathrm i\phi_i}),\qquad \phi_i=k_z^{(i)}d_i,\\
	\mathbf M&=\mathbf D_{0,1}\mathbf P_1\mathbf D_{1,2}\mathbf P_2\cdots\mathbf D_{59,60}\mathbf P_{60}\mathbf D_{60,\mathrm{sub}}.
	\end{aligned}
	\end{equation}
	Here medium 0 and the substrate are vacuum. The matrices are multiplied in the written order, mapping substrate amplitudes to entrance amplitudes. The complex reflection coefficient is $r=M_{21}/M_{11}$ and $R=|r|^2$.
	
	\textbf{Optimization algorithm.} The objective function $f(d_1, d_2) = -R(d_1, d_2; \theta_{\mathrm{target}})$ is minimized via a two-stage procedure:
	\begin{enumerate}
		\item \emph{Local optimization} using the Nelder--Mead simplex algorithm, initialized near the first-order Bragg condition: $d_1 \, \mathrm{Re}\left(\eps_1 - \sin^2\theta\right)^{1/2} + d_2 \, \mathrm{Re}\left(\eps_2 - \sin^2\theta\right)^{1/2} = \lambda/2$.
		\item \emph{Global optimization} using the differential evolution algorithm (\texttt{differential\_evolution} in SciPy, population-size multiplier \texttt{popsize=15} (30 candidates for two variables), up to 80 iterations) within bounds $d_1 \in [0.3, 5]$~nm, $d_2 \in [0.3, 9]$~nm, followed by quasi-Newton refinement using the L-BFGS-B method.
	\end{enumerate}
	The parameters yielding the highest $R$ for the target angle are retained.

	\subsubsection{Configuration 1: $\lambda = 13.5$~nm, Mo/Si} The absorbing material is Mo ($\delta = 0.0763$, $\beta = 0.0064$) and the spacer is Si ($\delta = 0.0010$, $\beta = 0.0018$). To retain the angles and numerical aperture of Example 4 at this wavelength, the mask period is scaled in proportion to $\lambda$, giving $L_x=8\lambda/\sin6^\circ\approx1033.2$~nm. At fixed $L_x\approx857$~nm the maximum wafer index would instead be $N=15$. Optimization results for the 38 tabulated facets are summarized in Table~\ref{tab:bragg_13_5nm_MoSi}.
	
	{\small
		\begin{longtable}{cccccccc}
			\caption{Layer thicknesses of optimized Bragg mirrors for Example 4 ($\lambda=13.5$~nm, Mo/Si, 60 layers, TE polarization). $d_{\mathrm{Mo}}$ and $d_{\mathrm{Si}}$ are layer thicknesses, $r$ is complex field reflection coefficient, $R=|r|^2$ is power reflectance at target incidence angle $\theta_{\mathrm{inc}}$.}\label{tab:bragg_13_5nm_MoSi}\\
			\toprule
			Mirror & $m$ & $\theta_{\mathrm{inc}}$ (${}^\circ$) & $d_{\mathrm{Mo}}$~(nm) & $d_{\mathrm{Si}}$~(nm) & $\operatorname{Re} r$ & $\operatorname{Im} r$ & $R = |r|^2$\\
			\midrule
			\endfirsthead
			\multicolumn{8}{c}{\tablename~\thetable~(Continued)}\\
			\toprule
			Mirror & $m$ & $\theta_{\mathrm{inc}}$ (${}^\circ$) & $d_{\mathrm{Mo}}$~(nm) & $d_{\mathrm{Si}}$~(nm) & $\operatorname{Re} r$ & $\operatorname{Im} r$ & $R = |r|^2$\\
			\midrule
			\endhead
			\bottomrule
			\endfoot
			\bottomrule
			\endlastfoot
			1 &  1 &  1.50 & 2.9983 & 3.9348 & 0.710212 & 0.461336 & 0.717232\\
			1 &  2 &  3.00 & 3.0000 & 3.9404 & 0.709757 & 0.462358 & 0.717530\\
			1 &  3 &  4.52 & 3.0029 & 3.9500 & 0.708982 & 0.464086 & 0.718032\\
			1 &  4 &  6.06 & 3.0071 & 3.9636 & 0.707865 & 0.466549 & 0.718741\\
			1 &  5 &  7.62 & 3.0125 & 3.9816 & 0.706379 & 0.469771 & 0.719656\\
			1 &  6 &  9.22 & 3.0194 & 4.0045 & 0.704467 & 0.473833 & 0.720792\\
			1 &  7 & 10.86 & 3.0278 & 4.0328 & 0.702079 & 0.478784 & 0.722149\\
			1 &  8 & 12.55 & 3.0380 & 4.0671 & 0.699135 & 0.484712 & 0.723736\\
			1 &  9 & 14.32 & 3.0503 & 4.1089 & 0.695495 & 0.491802 & 0.725582\\
			1 & 10 & 16.16 & 3.0650 & 4.1590 & 0.691065 & 0.500102 & 0.727673\\
			1 & 11 & 18.11 & 3.0826 & 4.2197 & 0.685606 & 0.509892 & 0.730045\\
			1 & 12 & 20.18 & 3.1039 & 4.2933 & 0.678903 & 0.521333 & 0.732697\\
			1 & 13 & 22.41 & 3.1300 & 4.3837 & 0.670585 & 0.534758 & 0.735651\\
			1 & 14 & 24.85 & 3.1628 & 4.4967 & 0.660149 & 0.550571 & 0.738925\\
			1 & 15 & 27.57 & 3.2056 & 4.6410 & 0.646893 & 0.569262 & 0.742530\\
			1 & 16 & 30.67 & 3.2643 & 4.8311 & 0.629874 & 0.591367 & 0.746457\\
			1 & 17 & 34.38 & 3.3532 & 5.0975 & 0.607430 & 0.617873 & 0.750738\\
			1 & 18 & 39.19 & 3.5124 & 5.5133 & 0.576892 & 0.650115 & 0.755454\\
			1 & 19 & 47.79 & 4.0065 & 6.5082 & 0.525839 & 0.696983 & 0.762293\\
			2 &  1 &  0.37 & 2.9978 & 3.9331 & 0.710353 & 0.461016 & 0.717138\\
			2 &  2 &  0.75 & 2.9979 & 3.9334 & 0.710325 & 0.461081 & 0.717157\\
			2 &  3 &  1.11 & 2.9980 & 3.9340 & 0.710280 & 0.461182 & 0.717186\\
			2 &  4 &  1.48 & 2.9983 & 3.9348 & 0.710216 & 0.461327 & 0.717229\\
			2 &  5 &  1.83 & 2.9986 & 3.9357 & 0.710138 & 0.461503 & 0.717280\\
			2 &  6 &  2.17 & 2.9989 & 3.9369 & 0.710046 & 0.461708 & 0.717340\\
			2 &  7 &  2.50 & 2.9993 & 3.9381 & 0.709943 & 0.461941 & 0.717409\\
			2 &  8 &  2.80 & 2.9997 & 3.9395 & 0.709836 & 0.462182 & 0.717479\\
			2 &  9 &  3.09 & 3.0002 & 3.9409 & 0.709720 & 0.462440 & 0.717554\\
			2 & 10 &  3.35 & 3.0006 & 3.9423 & 0.709607 & 0.462694 & 0.717628\\
			2 & 11 &  3.57 & 3.0010 & 3.9436 & 0.709504 & 0.462924 & 0.717695\\
			2 & 12 &  3.75 & 3.0013 & 3.9447 & 0.709415 & 0.463124 & 0.717753\\
			2 & 13 &  3.88 & 3.0016 & 3.9455 & 0.709348 & 0.463273 & 0.717796\\
			2 & 14 &  3.93 & 3.0017 & 3.9458 & 0.709321 & 0.463332 & 0.717814\\
			2 & 15 &  3.90 & 3.0016 & 3.9456 & 0.709337 & 0.463297 & 0.717803\\
			2 & 16 &  3.73 & 3.0013 & 3.9445 & 0.709425 & 0.463100 & 0.717746\\
			2 & 17 &  3.38 & 3.0006 & 3.9425 & 0.709594 & 0.462724 & 0.717637\\
			2 & 18 &  2.70 & 2.9996 & 3.9390 & 0.709873 & 0.462099 & 0.717454\\
			2 & 19 &  1.01 & 2.9980 & 3.9338 & 0.710294 & 0.461150 & 0.717177\\
		\end{longtable}
	}
	
	For the first mirror, the Mo layer thickness varies from $3.00$~nm (small angles) to $4.01$~nm ($m=19$, $\theta = 47.8^\circ$), while the Si layer thickness varies from $3.93$ to $6.51$~nm. The bilayer period $d = d_{\mathrm{Mo}} + d_{\mathrm{Si}}$ grows from $6.93$~nm at normal incidence to $10.51$~nm at $\theta = 47.8^\circ$, following the trend of the approximate Bragg condition $2d\cos\theta\approx\lambda$. Refraction is included through the layer-dependent expression used to initialize the optimization above. The reflectance $R$ rises from $0.717$ ($m=1$) to $0.762$ ($m=19$) --- as the angle of incidence increases, the effective penetration depth into the absorbing medium decreases, enhancing $R$. For the second mirror, all incidence angles remain small ($0.37^\circ$--$3.93^\circ$), so optimal thicknesses are virtually constant ($d_{\mathrm{Mo}} \approx 3.00$~nm, $d_{\mathrm{Si}} \approx 3.94$~nm), and $R$ varies only in the fourth decimal place ($0.7171$--$0.7178$). The small variation suggests that a common second-mirror coating can be a useful approximation. Its performance must be evaluated separately from the individually optimized designs tabulated here.

	\subsubsection{Configuration 2: $\lambda = 11.2$~nm, Ru/Be} The absorbing material is Ru ($\delta = 0.0660$, $\beta = 0.0065$) and the spacer is Be ($\delta = -0.0122$, $\beta = 0.0013$). A negative decrement indicates that the real part of the refractive index of Be at this wavelength exceeds unity, which occurs near the Be $K$-absorption edge at $\lambda \approx 11.3$~nm. Results are summarized in Table~\ref{tab:bragg_11_2nm_RuBe}.
	
	{\small
		\begin{longtable}{cccccccc}
			\caption{Layer thicknesses of optimized Bragg mirrors for Example 4 ($\lambda=11.2$~nm, Ru/Be, 60 layers, TE polarization). Notation as in Table~\ref{tab:bragg_13_5nm_MoSi}.}\label{tab:bragg_11_2nm_RuBe}\\
			\toprule
			Mirror & $m$ & $\theta_{\mathrm{inc}}$ (${}^\circ$) & $d_{\mathrm{Ru}}$~(nm) & $d_{\mathrm{Be}}$~(nm) & $\operatorname{Re} r$ & $\operatorname{Im} r$ & $R = |r|^2$\\
			\midrule
			\endfirsthead
			\multicolumn{8}{c}{\tablename~\thetable~(Continued)}\\
			\toprule
			Mirror & $m$ & $\theta_{\mathrm{inc}}$ (${}^\circ$) & $d_{\mathrm{Ru}}$~(nm) & $d_{\mathrm{Be}}$~(nm) & $\operatorname{Re} r$ & $\operatorname{Im} r$ & $R = |r|^2$\\
			\midrule
			\endhead
			\bottomrule
			\endfoot
			\bottomrule
			\endlastfoot
			1 &  1 &  1.50 & 2.3680 & 3.2983 & 0.669032 & 0.548699 & 0.748674\\
			1 &  2 &  3.00 & 2.3690 & 3.3030 & 0.668303 & 0.549863 & 0.748979\\
			1 &  3 &  4.52 & 2.3706 & 3.3112 & 0.667061 & 0.551835 & 0.749491\\
			1 &  4 &  6.06 & 2.3730 & 3.3228 & 0.665274 & 0.554640 & 0.750215\\
			1 &  5 &  7.62 & 2.3761 & 3.3382 & 0.662905 & 0.558308 & 0.751151\\
			1 &  6 &  9.22 & 2.3801 & 3.3577 & 0.659869 & 0.562927 & 0.752314\\
			1 &  7 & 10.86 & 2.3848 & 3.3817 & 0.656092 & 0.568551 & 0.753707\\
			1 &  8 & 12.55 & 2.3906 & 3.4110 & 0.651459 & 0.575275 & 0.755340\\
			1 &  9 & 14.32 & 2.3974 & 3.4466 & 0.645757 & 0.583304 & 0.757246\\
			1 & 10 & 16.16 & 2.4055 & 3.4893 & 0.638855 & 0.592685 & 0.759412\\
			1 & 11 & 18.11 & 2.4152 & 3.5411 & 0.630395 & 0.603726 & 0.761883\\
			1 & 12 & 20.18 & 2.4266 & 3.6039 & 0.620054 & 0.616599 & 0.764661\\
			1 & 13 & 22.41 & 2.4405 & 3.6810 & 0.607272 & 0.631665 & 0.767780\\
			1 & 14 & 24.85 & 2.4578 & 3.7773 & 0.591268 & 0.649364 & 0.771271\\
			1 & 15 & 27.57 & 2.4799 & 3.9005 & 0.570921 & 0.670234 & 0.775164\\
			1 & 16 & 30.67 & 2.5099 & 4.0629 & 0.544631 & 0.694871 & 0.779469\\
			1 & 17 & 34.38 & 2.5550 & 4.2910 & 0.509401 & 0.724400 & 0.784245\\
			1 & 18 & 39.19 & 2.6367 & 4.6471 & 0.459810 & 0.760344 & 0.789548\\
			1 & 19 & 47.79 & 2.9083 & 5.4854 & 0.373591 & 0.810425 & 0.796359\\
			2 &  1 &  0.37 & 2.3677 & 3.2968 & 0.669260 & 0.548334 & 0.748579\\
			2 &  2 &  0.75 & 2.3677 & 3.2971 & 0.669214 & 0.548407 & 0.748598\\
			2 &  3 &  1.11 & 2.3678 & 3.2975 & 0.669142 & 0.548523 & 0.748628\\
			2 &  4 &  1.48 & 2.3680 & 3.2982 & 0.669039 & 0.548688 & 0.748672\\
			2 &  5 &  1.83 & 2.3681 & 3.2990 & 0.668914 & 0.548888 & 0.748724\\
			2 &  6 &  2.17 & 2.3683 & 3.3000 & 0.668767 & 0.549123 & 0.748786\\
			2 &  7 &  2.50 & 2.3686 & 3.3011 & 0.668601 & 0.549389 & 0.748855\\
			2 &  8 &  2.80 & 2.3688 & 3.3022 & 0.668429 & 0.549663 & 0.748927\\
			2 &  9 &  3.09 & 2.3691 & 3.3034 & 0.668244 & 0.549958 & 0.749003\\
			2 & 10 &  3.35 & 2.3693 & 3.3046 & 0.668062 & 0.550247 & 0.749079\\
			2 & 11 &  3.57 & 2.3695 & 3.3057 & 0.667897 & 0.550509 & 0.749147\\
			2 & 12 &  3.75 & 2.3697 & 3.3066 & 0.667754 & 0.550736 & 0.749206\\
			2 & 13 &  3.88 & 2.3699 & 3.3073 & 0.667646 & 0.550908 & 0.749251\\
			2 & 14 &  3.93 & 2.3699 & 3.3076 & 0.667604 & 0.550975 & 0.749268\\
			2 & 15 &  3.90 & 2.3699 & 3.3075 & 0.667630 & 0.550934 & 0.749258\\
			2 & 16 &  3.73 & 2.3697 & 3.3065 & 0.667770 & 0.550711 & 0.749199\\
			2 & 17 &  3.38 & 2.3693 & 3.3048 & 0.668040 & 0.550282 & 0.749088\\
			2 & 18 &  2.70 & 2.3687 & 3.3018 & 0.668488 & 0.549568 & 0.748902\\
			2 & 19 &  1.01 & 2.3678 & 3.2974 & 0.669165 & 0.548486 & 0.748619\\
		\end{longtable}
	}
	
	The Ru/Be material combination at $\lambda = 11.2$~nm provides markedly higher reflectance than Mo/Si at $\lambda = 13.5$~nm. $R$ on the first mirror ranges from $0.749$ ($m=1$) to $0.796$ ($m=19$), which is $3$--$4$ percentage points higher than for Mo/Si ($0.717$--$0.762$). This is attributable to the lower absorption of Be ($\beta_{\mathrm{Be}} = 0.0013$) relative to Si ($\beta_{\mathrm{Si}} = 0.0018$) and the larger refractive index contrast of Ru/Be. As with Mo/Si, for the second mirror (small angles $0.37^\circ$--$3.93^\circ$) optimal layer thicknesses are practically constant: $d_{\mathrm{Ru}} \approx 2.37$~nm, $d_{\mathrm{Be}} \approx 3.30$~nm, $R \approx 0.749$.

	\subsubsection{Discussion} Maximizing reflectance over layer thicknesses at a fixed target angle does not force the maximum of the angular reflectance curve to occur at that angle. For example, for $m=1$ on the second mirror, the target is $0.37^\circ$, whereas the local near-normal maxima of the tabulated Mo/Si and Ru/Be designs occur at approximately $1.90^\circ$ and $2.00^\circ$, respectively. The tabulated reflectances are evaluated at the actual target angles. The angular-peak offset is not an optimization constraint.
	
	Comparison with the typical reflectance $R \sim 70\%$ of standard Mo/Si coatings (50 bilayers, $\lambda = 13.5$~nm, incident angle $\sim 6^\circ$) indicates that individually optimized 30-bilayer coatings achieve comparable or superior reflectance ($R = 71.7\%$--$76.2\%$) with fewer layers in the ideal model. This comparison is not an experimental coating-performance validation, because the calculation omits interface roughness, interdiffusion, and fabrication errors. For Ru/Be at $\lambda = 11.2$~nm, reflectance is even higher ($74.9\%$--$79.6\%$). The fraction of the power leaving the mask in an accepted order that is retained after the two modeled reflections is $R_1 R_2 \approx 51.5\%$ (Mo/Si, low angles) to $59.6\%$ (Ru/Be, $m=19$), exceeding the throughput of 6- and 10-mirror systems ($12\%$ and $2.8\%$, respectively) by up to a factor of $\sim 20$.

	\subsection{Mask Optimization for a Target Aerial Image on the Wafer}
	
	This section is dedicated to mask optimization for a prescribed field profile on the wafer in the synthesized projection system. Mask optimization is conducted using approaches developed in~\cite{eskin2026gradientbasedinverselithographyeuv}.

	\subsubsection{Formulation of the Inverse Problem}
	
	In Example 4, the optical path lengths of the channels with $1\le |n|\le N=19$ are equalized ($P_n\approx2998.1$~mm), removing order-dependent propagation phase shifts. The relative phases at the wafer still depend on the mask amplitudes and on the complex reflection coefficients of both Bragg coatings. Let $A_n$ denote the complex electric-field amplitude of mask order $n$, and let $r_{1,n}$ and $r_{2,n}$ denote the corresponding facet reflection coefficients. With the common propagation phase omitted and amplitudes referenced to $\zeta_{\mathrm{w}}=0$, the wafer coefficient is
	\begin{equation}
		B_{-n} = r_{2,n} r_{1,n} A_n.\label{eq:B_from_A}
	\end{equation}
	Coefficients $r_{1,n}$ and $r_{2,n}$ were determined previously (Table~\ref{tab:bragg_11_2nm_RuBe}) for $\lambda = 11.2$~nm, Ru/Be. The field on the wafer is evaluated via Eq.~\eqref{eq3}. Let $\mathbf{E}^{(d)}$ and $I^{(d)} = |\mathbf{E}^{(d)}|^2$ denote the target electric field distribution and intensity on the wafer. The objective of Inverse Lithography Technology (ILT) is to determine the spatial permittivity distribution $\eps_j(x)$ of the mask layers such that the projected aerial intensity matches the target intensity:
	\begin{equation}
		\left|\mathbf{E}^{(\mathrm{w})}\right|^2 = \left|\mathbf{E}^{(d)}\right|^2.
	\end{equation}
	We consider a mask comprising a single absorber layer with permittivity $\eps(x)$ positioned atop multilayer mirror layers that are homogeneous along $x$.

	\subsubsection{Forward Problem}
	
	Scattering amplitudes $A_n$ for a given profile $\eps(x)$ are evaluated using the modal waveguide method (equivalent to RCWA). The mask consists of an absorber layer of thickness $d_{\mathrm{abs}}$ with periodically modulated permittivity $\eps(x) = \sum_m \eps_m \exp(-\mathrm{i}\kappa_x m x)$ (where $\eps_m$ are Fourier coefficients), situated on a Bragg mirror (Ru/Be/Sr, 30 periods~\cite{eskin2026physicsinformedneuralsystemssimulation}). The solution proceeds as follows.
	
	The electric field inside the absorber layer is represented as $E_y(x,z) = X(x) Z(z)$. The transverse function $X(x)$ is expanded in a Fourier series over plane waves $\psi_m = \exp(-\mathrm{i}\kappa_x m x)$, and the longitudinal function is $Z(z) = \exp(\pm\mathrm{i} k_z z)$.
	
	Substituting into the Helmholtz equation yields the algebraic eigenvalue problem:
	\begin{equation}
		\hat{D}\mathbf{B}_p = k_{z;p}^2 \mathbf{B}_p, \qquad
		\hat{D}_{nm} = k_0^2\eps_{n-m} - (\kappa_x n)^2\delta_{nm},
		\label{eq:rcwa_eigen}
	\end{equation}
	where $k_{z;p}^2$ are the eigenvalues and $k_{z;p}$ are the longitudinal wavenumbers of the modes, and $\mathbf{B}_p$ are the eigenvectors (components $B_{p,m}$ define the contribution of the $m$-th plane wave to the $p$-th mode). Matrix $\hat{D}$ has dimensions $(2M+1)\times(2M+1)$.
	
	The field in each layer is expressed via modal expansions:
	\begin{equation}
		E_y^{(j)}(x,z) = k_0 \sum_{p=-M}^{M} \Bigl[A_{p;1}^{(j)}\mathrm{e}^{\mathrm{i}k_{z;p}^{(j)} z} + A_{p;2}^{(j)}\mathrm{e}^{-\mathrm{i}k_{z;p}^{(j)} z}\Bigr] \sum_{m=-M}^{M} B_{p,m}^{(j)}\psi_m.
	\end{equation}
	Coefficients $A_{p;1}^{(j)}, A_{p;2}^{(j)}$ are determined from tangential field continuity ($E_y$ and $H_x$) across layer boundaries, yielding the linear system:
	\begin{equation}
		\hat{\mathbf{M}}\mathbf{X} = \mathbf{R},
		\label{eq:rcwa_system}
	\end{equation}
	where $\mathbf{X}$ is the vector of unknown modal coefficients (reflected waves and waves transmitted into and through the mirror), and $\mathbf{R}$ describes the incident wave. Solving Eq.~\eqref{eq:rcwa_system} yields modal coefficients that must be converted to electric-field amplitudes. In the implemented normalization, $A_n=-k_0X_n^{(r)}$ ($n=-M,\ldots,M$), where $X_n^{(r)}$ is the reflected-wave block of $\mathbf X$. The minus sign follows its polarization convention. Here and in Eq.~\eqref{eq:B_from_A}, $A_n$ denotes the converted electric-field amplitude, not the raw modal coefficient. The reflected field above the mask ($z > 0$) is:
	\begin{equation}
		E_y^{(r)}(x,z) = \sum_{n=-M}^{M} A_n \exp(-\mathrm{i}\kappa_x n x - \mathrm{i} k_{z;n} z).
	\end{equation}

	\subsection{Optimization via Fourier-Parameterized Projection}
	
	To solve the inverse problem of finding $\eps(x)$ from the target wafer aerial image, we apply gradient-based optimization with Fourier-parameterized mask projection~\cite{eskin2026gradientbasedinverselithographyeuv}.
	
	Instead of optimizing hundreds of discrete pixels, a smooth latent function is introduced:
	\begin{equation}
		f_{\mathrm{latent}}(x) = \mathrm{Re}\left(\sum_{m=-M}^{M} F_m \mathrm{e}^{-\mathrm{i}\kappa_x m x}\right),
		\label{eq:latent}
	\end{equation}
	where $F_m$ are the complex Fourier coefficients to be optimized. The physical absorber density is obtained via a steepened sigmoid mapping:
	\begin{equation}
		\rho(x) = \sigma\bigl(\beta f_{\mathrm{latent}}(x)\bigr) \in (0, 1),
	\end{equation}
	where $\beta$ is a steepness parameter increased during optimization to enforce binarization. The permittivity is computed as:
	\begin{equation}
		\eps(x) = \eps_{\mathrm{vac}} + \rho(x)(\eps_{\mathrm{abs}} - \eps_{\mathrm{vac}}),
	\end{equation}
	where $\eps_{\mathrm{vac}} = 1$ and $\eps_{\mathrm{abs}}$ is the absorber permittivity. The nonlinearity of $\sigma$ allows a finite set of $F_m$ to generate an infinite high-frequency spectrum for $\eps(x)$, suppressing the Gibbs phenomenon.
	
	The inverse problem is to minimize the loss functional measuring the discrepancy between the intensities of the reflected and target fields on the wafer:
	\begin{equation}
		\eps(x) = \arg\min_{\eps} \mathcal{L}(\eps),
		\label{eq:opt_coord}
	\end{equation}
	where the total loss is:
	\begin{equation}
		\mathcal{L} = \mathcal{L}_{\mathrm{shape}} + \mathcal{L}_{\mathrm{bin}} + \mathcal{L}_{\mathrm{HF}} + \mathcal{L}_{\mathrm{supp}}.
		\label{eq:loss}
	\end{equation}
	Here:
	\begin{itemize}
		\item $\mathcal{L}_{\mathrm{shape}} = \mathrm{MSE}(I_{\mathrm{w}}^{\mathrm{norm}}, I_{\mathrm{target}}^{\mathrm{norm}})$ is the mean squared error between normalized wafer intensity and target profile, where $I_{\mathrm{w}}(x) = |E_y^{(\mathrm{w})}(x)|^2$ with $B_{-n} = r_{2,n} r_{1,n} A_n$.
		\item $\mathcal{L}_{\mathrm{bin}} = \langle \rho(1-\rho)\rangle$ penalizes intermediate grayscale density values, promoting binary solutions ($\rho \to 0$ or $1$).
		\item $\mathcal{L}_{\mathrm{HF}} = w_{\mathrm{HF}} \sum_{m=1}^{N_{\mathrm{FR}}} m |F_m|^2$ penalizes unphysical high-frequency spatial components, discouraging fine-scale structure without imposing a minimum manufacturable feature size (where $N_{\mathrm{FR}} = M$ is the Fourier truncation order).
		\item $\mathcal{L}_{\mathrm{supp}} = w_{\mathrm{supp}}(\alpha) |A_{-\tilde{m}}|^2$ suppresses scattering into the unused back-reflection order $-\tilde{m}$ that coincides with the input illumination aperture (this term is not present in~\cite{eskin2026gradientbasedinverselithographyeuv}).
	\end{itemize}
	
	In our experiments, the target intensity $I^{(d)}$ is binary, with unity in target regions and zero elsewhere. In the global wafer coordinates we prescribe $E_y^{(d)}(x)=\sqrt{I^{(d)}(x)}\exp(+4\mathrm{i}k_{0x}x)$, where $k_{0x}=k_0\sin6^\circ=\tilde m\kappa_x$. The numerical plots use the reversed wafer coordinate $x_{\mathrm p}=-x$ (and $y_{\mathrm p}=-y$ in 3D), so the implemented phase is $\exp(-4\mathrm{i}k_{0x}x_{\mathrm p})$. This convention accounts for the inversion of the transverse wavevector by the projection geometry. Decomposing the target field into spatial harmonics and retaining the selected wafer orders yields the bandlimited target intensity $\tilde I^{(d)}$ used in Eq.~\eqref{eq:loss}. The imposed phase is a modeling choice that affects this filtered target.
	
	The optimization procedure runs as follows: for the first $40\%$ of epochs, only the image shape is optimized ($\mathcal{L}_{\mathrm{bin}} = 0, \beta = 1$). Subsequently, $\beta$ is annealed up to $\sim 15$ while the weight of $\mathcal{L}_{\mathrm{bin}}$ increases to $5$, driving the mask toward a strict binary profile without violating physical admissibility. Optimization starts from an unconstrained latent field and converges to a physical density $\rho(x)$ that, after hard thresholding at $0.5$, defines the synthesized mask layout.
	
	We use the Adam optimizer with a learning rate of $5 \times 10^{-3}$ for 500 epochs. Gradients are computed via automatic differentiation through the full physical solver: $\partial\mathcal{L}/\partial{\bm \Theta} = (\partial\mathcal{L}/\partial\mathbf{E}^{(r)}) (\partial\mathbf{E}^{(r)}/\partial\eps) (\partial\eps/\partial{\bm \Theta})$, where ${\bm \Theta} = \{F_m\}$.
	
	Optimization was conducted for $\lambda = 11.2$~nm. Mask and projection parameters:
	\begin{itemize}
		\item Wavelength $\lambda = 11.2$~nm, incidence angle $6^\circ$, TE polarization.
		\item Mask period $L_x \approx 857$~nm, wafer period $L_x^{(\mathrm{w})} = L_x/4 \approx 214$~nm.
		\item Maximum harmonic index on the mask $M = 25$, on the wafer $N = 19$.
		\item Mask absorber: La ($\delta = -0.0440$, $\beta = 0.0159$) with thickness $d_{\mathrm{abs}} = 60$~nm.
		\item Bragg mirror beneath the absorber: 30 periods of Ru/Be/Sr ($d_{\mathrm{Ru}} = 1.7$~nm, $d_{\mathrm{Be}} = 2.7$~nm, $d_{\mathrm{Sr}} = 1.34$~nm).
		\item Mirror reflection coefficients $r_{1,n}, r_{2,n}$ taken from Table~\ref{tab:bragg_11_2nm_RuBe} (Ru/Be, 30 bilayers).
		\item Target profile on the wafer: intensity maximum at $x = 0$ within the cell $[-107, 107]$~nm.
		\item Fourier expansion parameter $M_{\mathrm{opt}} = M = 25$ (the truncation order of the latent-function expansion, $N_{\mathrm{FR}}$ in Eq.~\eqref{eq:loss}), 500 epochs.
	\end{itemize}

	\subsubsection{Results of Optimizations}
	
	The unfiltered target intensity is non-zero over $x\in[-2.3,2.3]$~nm, corresponding to a rectangular width of $4.6$~nm. The full width at half maximum (FWHM) estimated from the plotted bandlimited target is approximately $5.1$~nm, whereas that of the synthesized central peak is approximately $5.4$~nm. These widths differ from the unfiltered target width and are not fixed a priori to $\lambda/2$.
	
\begin{figure}[ht!]\centering
		\includegraphics[width=0.5\textwidth]{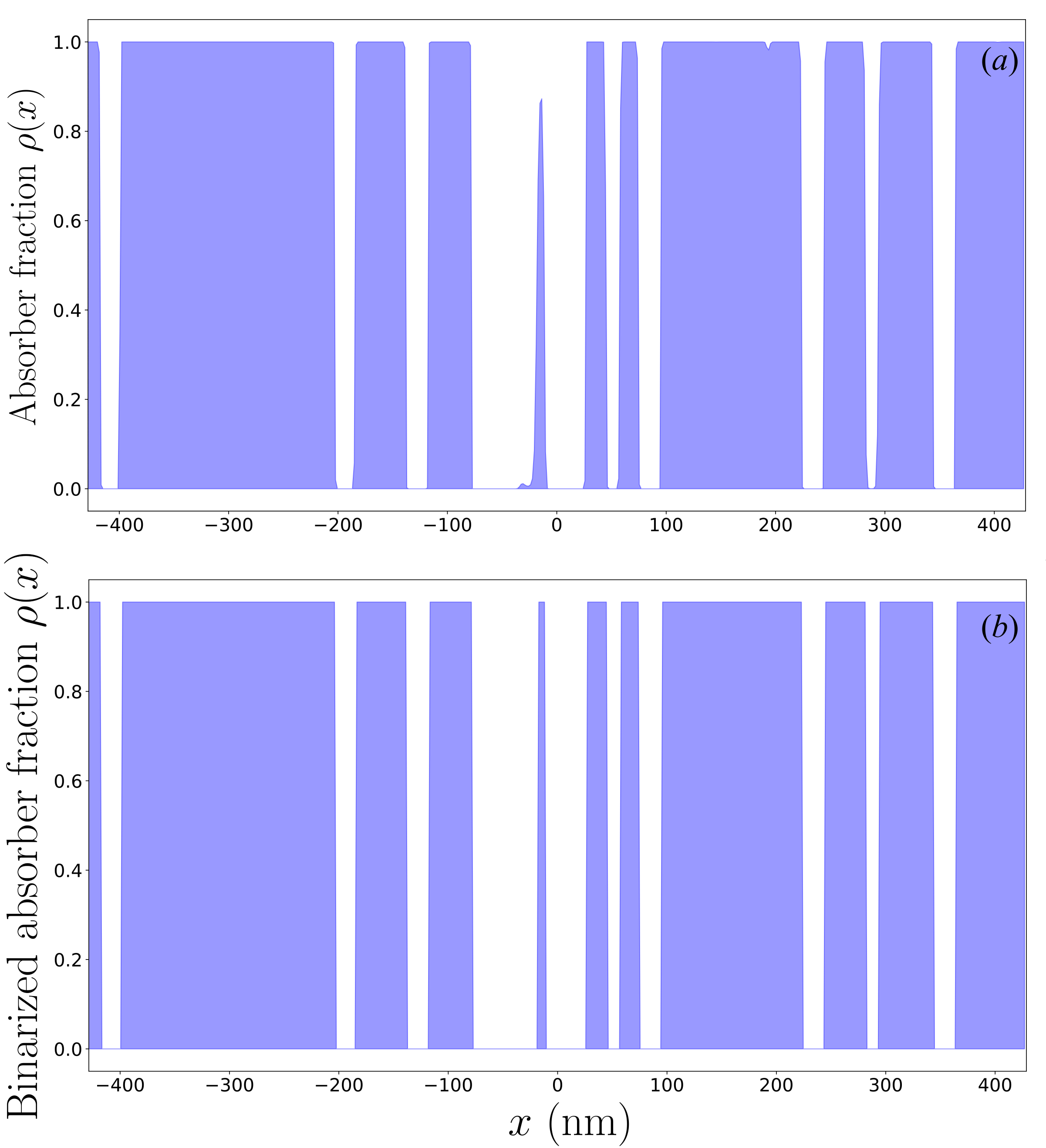}
		\caption{(a) Absorber density distribution $\rho(x)$ in the optimized mask ($\lambda=11.2$~nm, La absorber, 500 epochs). (b) Binarized mask profile ($\rho \in \{0, 1\}$ at threshold $\rho > 0.5$).}\label{fig:mask_profile}
	\end{figure}

	Figure~\ref{fig:mask_profile}a illustrates the optimized continuous absorber density $\rho(x)$. In Fig.~\ref{fig:mask_profile}b, the binarized mask layout ($\rho(x) \in \{0, 1\}$) is shown. The mask period $L_x \approx 857$~nm contains several absorber features synthesized to shape the required diffraction spectrum, whose fabrication by electron-beam lithography would require separate verification of minimum widths, gaps, and fabrication tolerances.
	
\begin{figure}[ht!]\centering
		\includegraphics[width=0.5\textwidth]{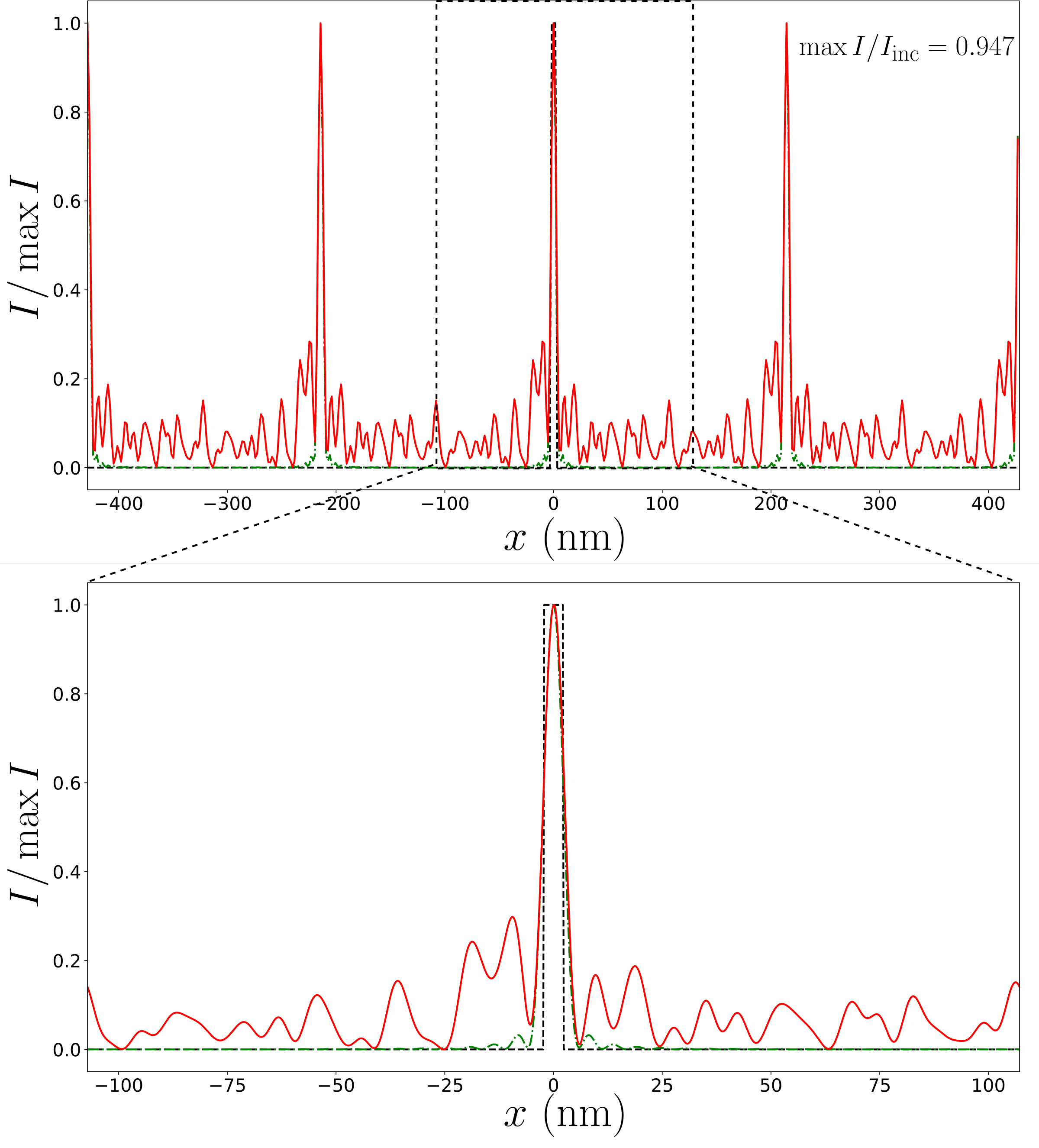}
		\caption{Aerial intensity on the wafer across $[-L_x/2, L_x/2]$: target profile (dashed line), bandlimited target with propagating orders (dash-dotted line), and optimized result (solid line). The peak intensity ratio is $\max I / I_{\mathrm{inc}} = 0.947$. Periodicity $L_x^{(\mathrm{w})} = 214$~nm.}\label{fig:wafer_intensity}
	\end{figure}

	Figure~\ref{fig:wafer_intensity} compares the target aerial image with the optimized result across $[-L_x/2, L_x/2] = [-428.6, 428.6]$~nm. The field repeats with period $L_x^{(\mathrm{w})} = 214$~nm. The peak profile is reproduced with low error ($\mathcal{L}_{\mathrm{shape}} = 1.6 \times 10^{-5}$). The peak intensity ratio reaches $\max I / I_{\mathrm{inc}} = 0.947$ ($94.7\%$ of incident intensity), which exceeds the baseline two-reflection reflectance $R^2 \approx (0.749)^2 \approx 56\%$ due to constructive interference among the diffracted orders.
	
\begin{figure}[ht!]\centering
		\includegraphics[width=\textwidth]{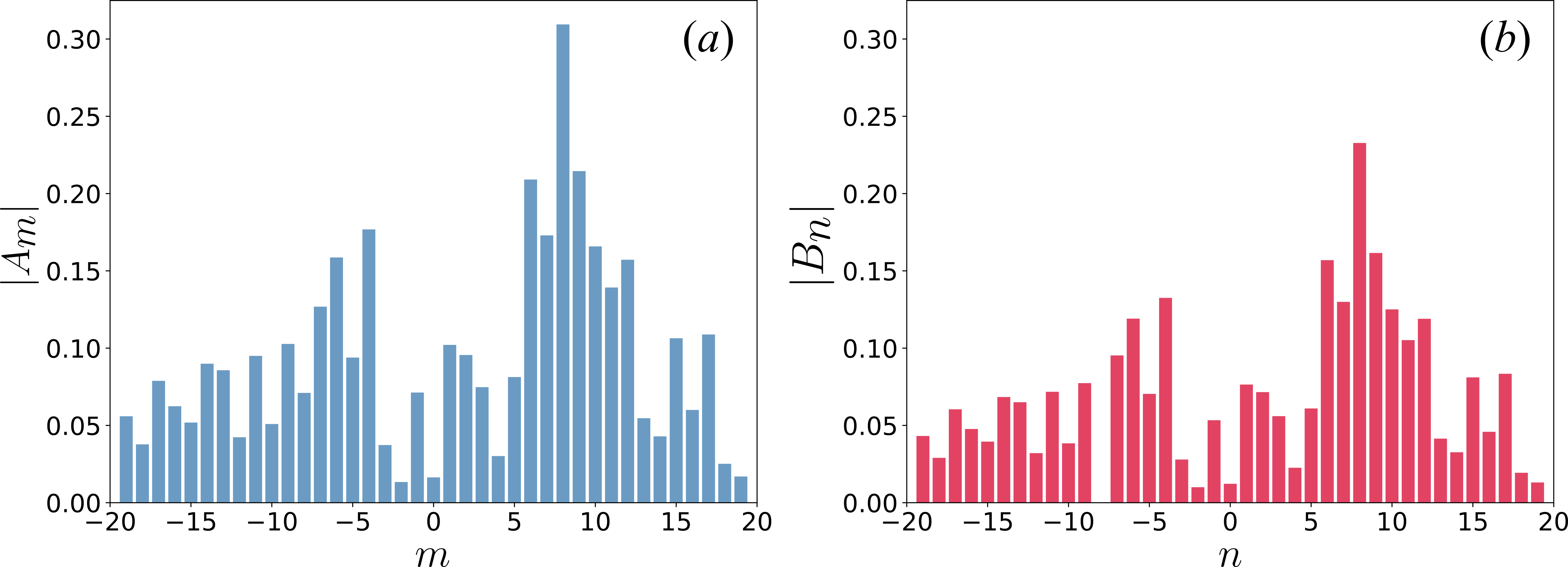}
		\caption{Field amplitude spectra: (a) is $|A_m|$ scattered by the mask, (b) is the wafer spectrum indexed by the originating mask order, corresponding to $|B_{-m}|$ in global coordinates.}\label{fig:spectrum_AB}
	\end{figure}

	Figure~\ref{fig:spectrum_AB} depicts the harmonic amplitude spectra $|A_m|$ (scattered by the mask) and $|B_{-m}| = |r_{2,m} r_{1,m} A_m|$ (incident on the wafer). The specular reflection order $m = 8$ (coinciding with the incident-beam order $\tilde{m} = 8$) dominates, while higher orders decay with $|m|$. The coefficients $B_{-m}$ are additionally modulated by $r_{1,m}$ and $r_{2,m}$ and largely follow the behavior of $|A_m|$. The plotted wafer spectrum is indexed by its originating mask order $m$, so its coefficient is $B_{-m}$ in the global-coordinate convention.
	
\begin{figure}[ht!]\centering
		\includegraphics[width=0.5\textwidth]{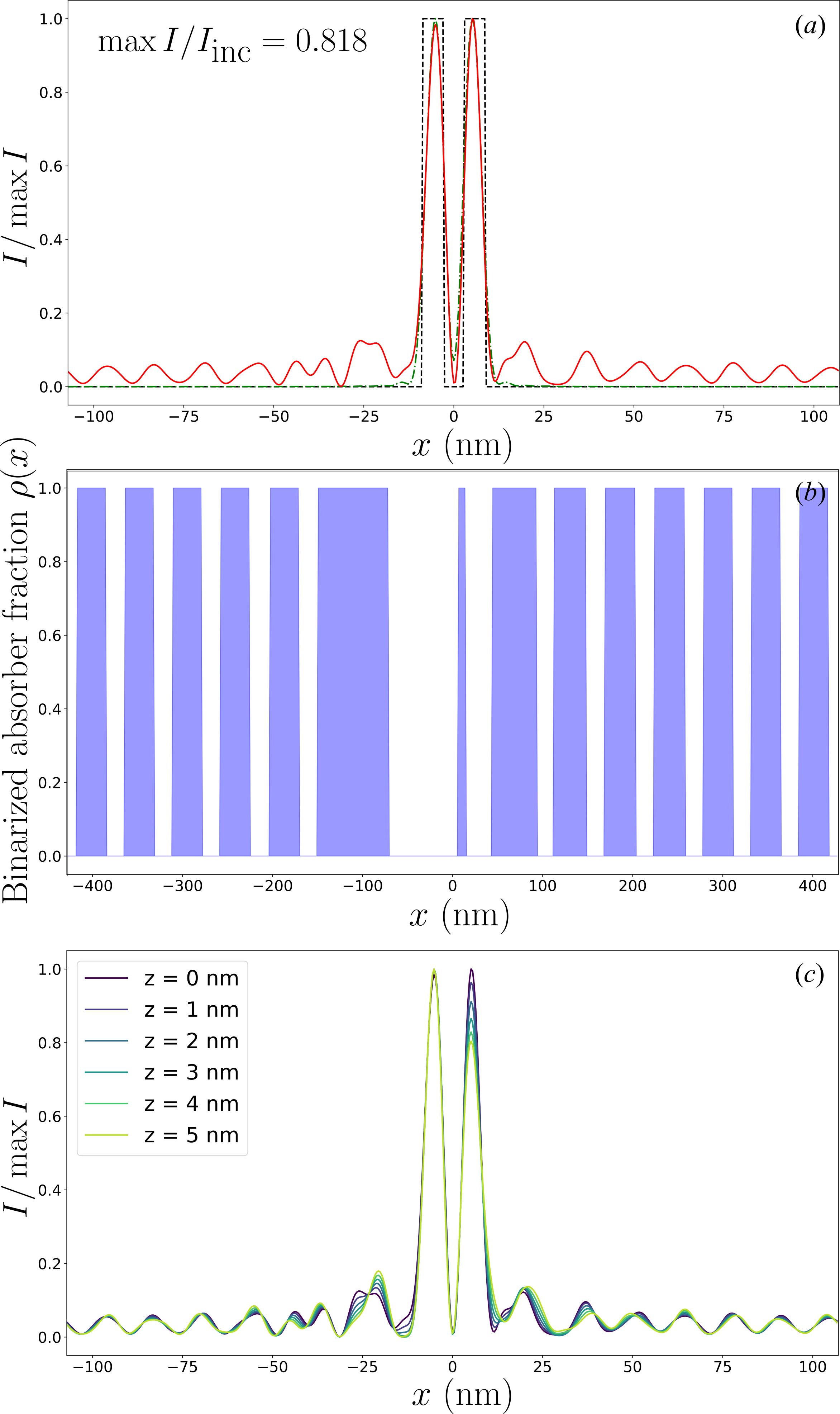}
		\caption{(a) Normalized aerial intensity on the wafer across $[-L_x/8, L_x/8]$: target profile (dashed), bandlimited target (dash-dotted), and optimized result (solid). The peak intensity ratio is $\max I / I_{\mathrm{inc}} = 0.818$. Periodicity $L_x^{(\mathrm{w})} = 214$~nm. (b) Synthesized binary mask profile ($\rho \in \{0, 1\}$). (c) Normalized wafer aerial intensity under wafer defocus along the $z$-axis for $\Delta z = 0, \ldots, 5$~nm.}\label{fig:wafer_stability}
	\end{figure}

	We next evaluated mask synthesis for an aerial image with two closely spaced intensity peaks in Example~4, targeting two $6$-nm-wide lines separated by a $6$-nm space within the periodic cell. Figure~\ref{fig:wafer_stability}(a,b) shows the simulated image and the synthesized binary mask. These are aerial-image calculations, not a resist-printing simulation.
	
	To evaluate sensitivity to axial wafer displacement, the field was computed for wafer displacements $\Delta z \in [0, 5]$~nm. As shown in Fig.~\ref{fig:wafer_stability}(c), the two peaks remain resolved for the tested defocus values. The calculation does not establish a complete focus--dose process window.
	
	\newpage
	\section{3D Two-Mirror Projection System}\label{3Dmirrors}
	
	\subsection{Problem Formulation and Basic Equations}
	
	The preceding analysis addressed a 2D formulation (TE polarization, $\mathbf{E} \parallel \mathbf{y}$, mask invariant along $y$). In three dimensions, the mask represents a 2D periodic structure with spatial periods $L_x$ along $x$ and $L_y$ along $y$ (assuming $L_x = L_y$, a square lattice). The mask occupies the domain $[-L/2, L/2] \times [-L/2, L/2] \times [-D, 0]$ in Cartesian coordinates $(x, y, z)$. A TE-polarized beam is incident at $6^\circ$ to the $z$-axis in the $xz$-plane: $\mathbf{k}_0 = (k_0\sin(\pi/30), 0, -k_0\cos(\pi/30))$ (the beam propagates toward the mask, in the $-z$ direction).
	
	The transverse component of the scattered electric field outside the mask is expanded in a double Fourier series:
	\begin{equation}
		\begin{bmatrix}
			E_x^{(r)}\\
			E_y^{(r)}
		\end{bmatrix} = \sum_{m=-\infty}^{\infty} \sum_{n=-\infty}^{\infty} \begin{bmatrix}
			 A_{x;m,n}^{(r)}\\
			A_{y;m,n}^{(r)}
		\end{bmatrix} \exp\bigl(-\mathrm{i}\kappa_x m x - \mathrm{i}\kappa_y n y - \mathrm{i} k_{z;m,n} z\bigr),
		\label{eq:refl_3d}
	\end{equation}
	where $A_{x;m,n}^{(r)}$ and $A_{y;m,n}^{(r)}$ are the amplitudes of the $(m,n)$-th harmonic, the superscript $(r)$ denotes the reflected (mask-side) field, $\kappa_x = 2\pi/L_x, \kappa_y = 2\pi/L_y$, and $k_{z;m,n} = \left(k_0^2 - \kappa_x^2 m^2 - \kappa_y^2 n^2\right)^{1/2}$ ($\mathrm{Im}\, k_{z;m,n} \le 0$). Propagating orders satisfy $\kappa_x^2 m^2 + \kappa_y^2 n^2 \le k_0^2$ ($M = \lfloor k_0/\kappa_x \rfloor$, the maximum order along each axis). The propagating field in the GO beam approximation is:
	\begin{equation}
		\begin{bmatrix}
			E_x^{(r)}\\
			E_y^{(r)}
		\end{bmatrix} = \sum_{\substack{m,n \\ \kappa_x^2 m^2 + \kappa_y^2 n^2 \le k_0^2}} \begin{bmatrix}
		\tilde{A}_{x;m,n}^{(r)}(x,y,z)\\
		\tilde{A}_{y;m,n}^{(r)}(x,y,z)
		\end{bmatrix} \exp\bigl(-\mathrm{i}\kappa_x m x - \mathrm{i}\kappa_y n y - \mathrm{i} k_{z;m,n} z\bigr),
		\label{eq2_3d}
	\end{equation}
	where $\tilde{A}_{x;m,n}^{(r)}$ and $\tilde{A}_{y;m,n}^{(r)}$ are the amplitudes within the beam of order $(m,n)$, and zero elsewhere.
	
	The field on the wafer under $4\times$ demagnification ($L_x^{(\mathrm{w})} = L_x/4, L_y^{(\mathrm{w})} = L_y/4$) is:
	\begin{equation}
		\begin{bmatrix}
			E_x^{(\mathrm{w})}\\
			E_y^{(\mathrm{w})}
		\end{bmatrix} = \sum_{(m,n)\in\mathcal{D}_N} \begin{bmatrix}
		B_{x;m,n}^{(\mathrm{w})}\\
		B_{y;m,n}^{(\mathrm{w})}
		\end{bmatrix}  \exp\bigl(-\mathrm{i}\kappa_x^{(\mathrm{w})} m x - \mathrm{i}\kappa_y^{(\mathrm{w})} n y - \mathrm{i}k_{z;m,n}^{(\mathrm{w})}\zeta_{\mathrm{w}} \bigr),
		\label{eq3_3d}
	\end{equation}
	where $k_{z;m,n}^{(\rm w)} = \bigl[k_0^2 - (\kappa_x^{(\rm w)} m)^2 - (\kappa_y^{(\rm w)} n)^2\bigr]^{1/2}$, $\kappa_x^{(\mathrm{w})} = 4\kappa_x, \kappa_y^{(\mathrm{w})} = 4\kappa_y$, and $(\mathrm{w})$ denotes the wafer-side field. Here $\zeta_{\mathrm{w}}=z-Z_1$, and the coefficients are referenced to the wafer plane. For the square lattice, harmonics propagating toward the wafer satisfy $m^2+n^2<[k_0/(4\kappa_x)]^2$. We select the smaller circular aperture $m^2+n^2\le N^2$, where $N=\lfloor k_0/(4\kappa_x)\rfloor=19$. This aperture is a design choice, not the exact propagation boundary. At the stated parameters, 24 additional propagating orders lie outside it. The selected domain is
	\begin{equation}
		\mathcal{D}_N=\{(m,n) \in \mathbb{Z}^2 : m^2 + n^2 \le N^2, \, (m,n) \ne (0,0)\}.
	\end{equation}
	
	Each $(m,n)$-th beam is characterized by polar angle $\theta_{m,n}$ (from the $z$-axis) and azimuthal angle $\varphi_{m,n}$ (in the $xy$-plane):
	\begin{equation}
		\sin\theta_{m,n} = \frac{\left[(\kappa_x m)^2 + (\kappa_y n)^2\right]^{1/2}}{k_0}, \qquad
		\varphi_{m,n} = {\rm Arg}(m + \mathrm{i} n).
		\label{eq:angles_3d}
	\end{equation}
	On the wafer, the transverse wavevector, and hence $\sin\theta$, is scaled by a factor of 4: $\sin\theta_{m,n}^{(\mathrm{w})} = 4\sin\theta_{m,n}$, with $\mathrm{NA}_{\max} = \sin\theta_{N,0}^{(\mathrm{w})} = 0.993$.
	
	The two-mirror system generalizes naturally to 3D. The first mirror is positioned in the plane $z = Z_1 = 1000$~mm, with the center of facet $(m,n)$ located at:
	\begin{equation}
		(x_1^{(m,n)},\, y_1^{(m,n)},\, Z_1) = \bigl(Z_1 \tan\theta_{m,n} \cos\varphi_{m,n},\; Z_1 \tan\theta_{m,n} \sin\varphi_{m,n},\; Z_1\bigr).
		\label{eq:mirror1_3d}
	\end{equation}
	The second mirror redirects the beam toward the wafer center $(0, 0, Z_1)$ at angle $\theta_{m,n}^{(\mathrm{w})}$. The facet center on the second mirror is positioned at:
	\begin{equation}
		(x_f^{(m,n)},\, y_f^{(m,n)},\, z_f^{(m,n)}) = \bigl(u_{m,n} \tan\theta_{m,n}^{(\mathrm{w})} \cos\varphi_{m,n},\; u_{m,n} \tan\theta_{m,n}^{(\mathrm{w})} \sin\varphi_{m,n},\; Z_1 - u_{m,n}\bigr),
		\label{eq:mirror2_3d}
	\end{equation}
	where $u_{m,n} = Z_1 - z_f^{(m,n)}$ enforces equalized path lengths. The deflection angle $\beta_{m,n}$ from the vertical satisfies:
	\begin{equation}
		\tan\beta_{m,n} = \frac{\left[(x_f - x_1)^2 + (y_f - y_1)^2\right]^{1/2}}{u_{m,n}}.
	\end{equation}
	The total optical path length is:
	\begin{equation}
		P_{m,n} = \frac{Z_1}{\cos\theta_{m,n}} + \left[(x_f - x_1)^2 + (y_f - y_1)^2 + u_{m,n}^2\right]^{1/2} + \frac{u_{m,n}}{\cos\theta_{m,n}^{(\mathrm{w})}}.
		\label{eq:path_3d}
	\end{equation}
	Parameter $u_{m,n}$ is determined by bisection to enforce $P_{m,n} \approx 2998.1$~mm. Incidence angles on the facets are $i_1 = (\theta_{m,n} + \beta_{m,n})/2$, while $i_2$ is computed via scalar products.
	
	For a square grating ($L_x = L_y$), all radial parameters ($\theta, \theta^{(\mathrm{w})}, z_f, \rho_f, \beta, i_1, i_2, P$), where $\rho_f=(x_f^2+y_f^2)^{1/2}$, depend solely on the radius $r = \left(m^2 + n^2\right)^{1/2}$ and are independent of $\varphi$. The facet coordinates $(x,y)$ are obtained via rotation by $\varphi$:
	\begin{equation}
		x_1 = Z_1 \tan\theta \cos\varphi, \quad y_1 = Z_1 \tan\theta \sin\varphi, \quad x_f = \rho_f(r)\cos\varphi, \quad y_f = \rho_f(r)\sin\varphi.
	\end{equation}
	Thus, the radial design reduces to the 2D problem (Example 4) by substituting $m \to r$. For non-integer $r$ (most orders with $n \ne 0$), the radial solution is obtained numerically by bisection, and the reflectances $R_1, R_2$ are interpolated linearly over $r$ from Table~\ref{tab:bragg_11_2nm_RuBe}.

	\subsection{Design of the 3D Two-Mirror Projection System}
	
	\subsubsection{Computational Parameters}
	
	Parameters match Example 4: $\lambda = 11.2$~nm, $L = 10$~mm, $L_x = L_y \approx 857$~nm, $Z_1 = 1000$~mm, $4\times$ demagnification, $M = 76, N = 19$, $\mathrm{NA}_{\max} = 0.993$, and total optical path length $2998.1$~mm. Harmonic $(m, n) = (-8, 0)$ (the back-reflection) does not contribute to the aerial image because it coincides with the illumination input aperture.
	
	The number of orders in the selected wafer aperture is 1128 (all integer pairs $(m,n)$ within $m^2 + n^2 \le 19^2$, excluding the origin). Each order maps to one facet on mirror 1 and one on mirror 2, yielding 2256 facets total. The number of unique $r$ values is 132. Due to rotational symmetry, each $r$ corresponds to 4 to 24 individual orders that share identical Bragg multilayer parameters.

	\subsubsection{Results for Facet Layouts}
	
	Table~\ref{tab:3d_orders} lists facet centers, incidence angles, and interpolated reflectances for 21 representative orders $(m, n)$. The total optical path is $P_{m,n} = 2998.1$~mm across all orders. Facet tilt $\alpha_1$ depends only on $r$ and ranges from $0.75^\circ$ ($r = 1$) to $33.4^\circ$ ($r = 19$). For the second mirror, $\alpha_2$ ranges from $2.6^\circ$ to $82.2^\circ$.
	
	{
		\begingroup\setlength{\tabcolsep}{3pt}
\begin{longtable}{cccccccccccc}
			\caption{Facet configurations of the first and second mirrors for the 3D case ($\lambda = 11.2$~nm, $L_x = L_y = 857$~nm, $N = 19$, $\mathrm{NA} = 0.993$, $P = 2998.1$~mm). $r = \left(m^2 + n^2\right)^{1/2}$, $\varphi$ is azimuthal angle, $(x_1, y_1)$ is facet center on mirror 1 at $z = Z_1 = 1000$~mm, $(x_f, y_f, z_f)$ is facet center on mirror 2, $i_1$ and $i_2$ are incidence angles, $|R_1 R_2|$ is product of reflectances (interpolated from Table~\ref{tab:bragg_11_2nm_RuBe}).}\label{tab:3d_orders}\\
			\toprule
			$m$ & $n$ & $r$ & $\varphi$ (${}^\circ$) & $x_1$ (mm) & $y_1$ (mm) & $x_f$ (mm) & $y_f$ (mm) & $z_f$ (mm) & $i_1$ (${}^\circ$) & $i_2$ (${}^\circ$) & $|R_1 R_2|$ \\
			\midrule
			\endfirsthead
			\multicolumn{12}{c}{\tablename~\thetable~(Continued)}\\
			\toprule
			$m$ & $n$ & $r$ & $\varphi$ (${}^\circ$) & $x_1$ (mm) & $y_1$ (mm) & $x_f$ (mm) & $y_f$ (mm) & $z_f$ (mm) & $i_1$ (${}^\circ$) & $i_2$ (${}^\circ$) & $|R_1 R_2|$ \\
			\midrule
			\endhead
			\bottomrule
			\endfoot
			\bottomrule
			\endlastfoot
			1 &  0 & 1.000 &  0.0 &  13.1 &   0.0 &   52.2 &    0.0 &    2.0 &  1.50 & 0.37 & 0.560\\
			1 &  1 & 1.414 & 45.0 &  13.1 &  13.1 &   52.2 &   52.2 &    3.2 &  2.12 & 0.53 & 0.561\\
			2 &  1 & 2.236 & 26.6 &  26.1 &  13.1 &  104.6 &   52.3 &    6.5 &  3.36 & 0.83 & 0.561\\
			2 &  2 & 2.828 & 45.0 &  26.1 &  26.1 &  104.6 &  104.6 &    9.9 &  4.26 & 1.05 & 0.561\\
			3 &  0 & 3.000 &  0.0 &  39.2 &   0.0 &  157.0 &    0.0 &   11.0 &  4.52 & 1.11 & 0.561\\
			3 &  2 & 3.606 & 33.7 &  39.2 &  26.2 &  157.2 &  104.8 &   15.6 &  5.45 & 1.33 & 0.561\\
			4 &  3 & 5.000 & 36.9 &  52.4 &  39.3 &  210.2 &  157.7 &   29.4 &  7.62 & 1.83 & 0.562\\
			5 &  0 & 5.000 &  0.0 &  65.5 &   0.0 &  262.8 &    0.0 &   29.4 &  7.62 & 1.83 & 0.562\\
			7 &  0 & 7.000 &  0.0 &  91.8 &   0.0 &  370.2 &    0.0 &   58.3 & 10.86 & 2.50 & 0.564\\
			5 &  5 & 7.071 & 45.0 &  65.6 &  65.6 &  264.5 &  264.5 &   59.5 & 10.98 & 2.52 & 0.565\\
			8 &  5 & 9.434 & 32.0 & 105.3 &  65.8 &  427.7 &  267.3 &  110.2 & 15.11 & 3.21 & 0.568\\
			7 &  7 & 9.899 & 45.0 &  92.2 &  92.2 &  375.1 &  375.1 &  122.6 & 15.97 & 3.32 & 0.569\\
			10 &  0 &10.000 &  0.0 & 131.8 &   0.0 &  536.2 &    0.0 &  125.4 & 16.16 & 3.35 & 0.569\\
			10 &  7 &12.207 & 35.0 & 132.4 &  92.6 &  543.5 &  380.5 &  199.1 & 20.63 & 3.78 & 0.573\\
			13 &  0 &13.000 &  0.0 & 172.4 &   0.0 &  710.7 &    0.0 &  232.5 & 22.41 & 3.88 & 0.575\\
			10 & 10 &14.142 & 45.0 & 133.0 & 133.0 &  551.6 &  551.6 &  289.1 & 25.22 & 3.93 & 0.578\\
			16 &  0 &16.000 &  0.0 & 213.8 &   0.0 &  897.5 &    0.0 &  411.4 & 30.67 & 3.73 & 0.584\\
			15 &  8 &17.000 & 28.1 & 201.0 & 107.2 &  850.1 &  453.4 &  502.4 & 34.38 & 3.38 & 0.588\\
			18 &  1 &18.028 &  3.2 & 242.0 &  13.4 & 1031.9 &   57.3 &  632.5 & 39.35 & 2.67 & 0.591\\
			13 & 13 &18.385 & 45.0 & 175.0 & 175.0 &  748.4 &  748.4 &  694.9 & 41.64 & 2.27 & 0.593\\
			19 &  0 &19.000 &  0.0 & 256.3 &   0.0 & 1102.1 &    0.0 &  869.1 & 47.79 & 1.01 & 0.596\\
		\end{longtable}
\endgroup
	}
	
\begin{figure}[ht!]\centering
		\includegraphics[width=\textwidth]{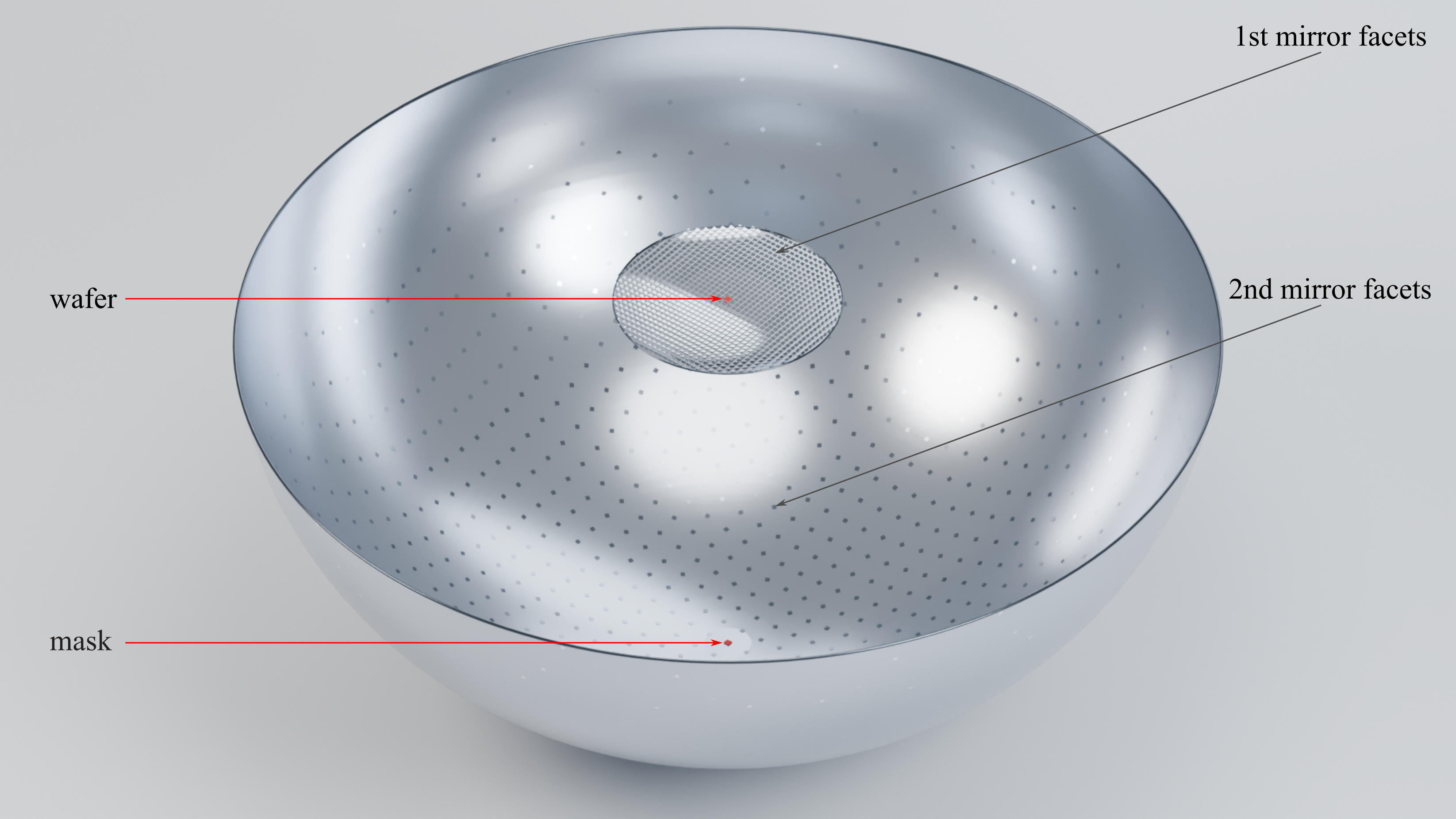}
		\caption{3D CAD visualization of the all-reflective two-mirror system (viewed at $60^\circ$ to the $z$-axis). Gray planar elements: 1128 facets of mirror 1 (in the plane $z = Z_1 = 1000$~mm, disk radius $256.3$~mm) and 1128 facets of mirror 2 (on the surface of revolution from $z_f = 2$~mm to $869$~mm). Red $1 \times 1$~cm squares denote the mask ($z=0$) and wafer ($z = Z_1$).}\label{fig:3d_render}
	\end{figure}

	The centers of the 1128 facets of the first mirror lie in a common plane at $z = Z_1 = 1000$~mm within a circle of radius $256.3$~mm. The 1128 facets of the second mirror lie on a surface of revolution spanning $z_f(r) = 2.0$~mm ($r = 1$) to $869.1$~mm ($r = 19$). Figure~\ref{fig:3d_render} shows a 3D rendering of the projection architecture. For clarity, the silver-colored backing of the second mirror is shown in the figure only to highlight the facets against its background. In the actual system, no such underlying surface exists.
	
	Angles of incidence on mirror 1 range from $i_1 = 1.50^\circ$ ($r = 1$) to $47.79^\circ$ ($r = 19$), while on mirror 2 they remain between $i_2 = 0.37^\circ$ and $3.93^\circ$. The two-reflection throughput $|R_1 R_2|$ varies narrowly between $0.560$ and $0.596$ (mean $0.576$), retaining $\sim 58\%$ of the power leaving the mask in each selected order after the two reflections (4--20 times more than in 6--10-mirror schemes).

	\textbf{Practical limitations of the 3D case.} The total of 2256 facets (1128 per mirror) makes direct fabrication and alignment technically challenging. However, due to rotational symmetry, 132 radial coating pairs are required, one per radial value $r$, corresponding to up to 264 distinct multilayer structures for the two mirrors rather than one design per facet. Replacing the second-mirror coatings by a common near-normal-incidence coating is a separate approximation.

	\subsection{3D Mask Synthesis}
	
	\subsubsection{Formulation of the 3D Inverse Problem}
	
	In 3D, all 1128 diffraction orders have equal optical path lengths: $P_{m,n} \approx 2998.1$~mm. The diffracted field contains two independent orthogonal transverse components, $E_x$ and $E_y$, corresponding to scattering amplitudes $A_{x;m,n}^{(r)}$ and $A_{y;m,n}^{(r)}$ computed via the 3D waveguide method.
	
	The harmonic amplitudes on the wafer are:
	\begin{equation}
		\begin{bmatrix}
			B_{x;-m,-n}^{(\mathrm{w})}\\
			B_{y;-m,-n}^{(\mathrm{w})}
		\end{bmatrix} 
		= \hat{r}_{2,m,n} \hat{r}_{1,m,n}  
		\begin{bmatrix}
			A_{x;m,n}^{(r)}\\
			A_{y;m,n}^{(r)}
		\end{bmatrix} , \qquad (m,n) \in \mathcal{D}_N,
		\label{eq:B_from_A_3d}
	\end{equation}
	where $\hat{r}_{1,m,n}$ and $\hat{r}_{2,m,n}$ denote the complex matrix field reflection coefficients of the corresponding facets of the first and second mirrors for order $(m,n)$, $\mathcal{D}_N = \{(m,n) \in \mathbb{Z}^2 : m^2+n^2 \le N^2, \, (m,n) \ne (0,0)\}$.

	For the general vector transfer, each matrix acts on global transverse electric-field components and includes the local polarization bases,
	\[
	\hat r_j=Q_j^{\mathrm{out}}\operatorname{diag}(r_{s,j},r_{p,j})(Q_j^{\mathrm{in}})^{-1}.
	\]
	The columns of $Q_j^{\mathrm{in}}$ and $Q_j^{\mathrm{out}}$ are the $xy$ projections of the incoming and outgoing local TE- and TM-polarization unit vectors, with the reflection coefficients defined in these same bases. The first reflection therefore acts before the second. The numerical demonstration retains the scalar-channel approximation with equal TE and TM coefficients and evaluates only transverse field components, as a proof-of-concept simplification.
	
	The field components above the wafer are:
	\begin{align}
		\begin{bmatrix}
			E_x^{(\mathrm{w})}(x,y)\\
			E_y^{(\mathrm{w})}(x,y)
		\end{bmatrix}  &= \sum_{(m,n) \in \mathcal{D}_N} \begin{bmatrix}
			B_{x;m,n}^{(\mathrm{w})}\\
			B_{y;m,n}^{(\mathrm{w})}
		\end{bmatrix}  \exp\bigl(-\mathrm{i}\kappa_x^{(\mathrm{w})} m x - \mathrm{i}\kappa_y^{(\mathrm{w})} n y\bigr),
		\label{eq:Ex_wafer_3d}
	\end{align}
	and the total aerial intensity is:
	\begin{equation}
		I_{\mathrm{w}}(x,y) = \bigl|E_x^{(\mathrm{w})}(x,y)\bigr|^2 + \bigl|E_y^{(\mathrm{w})}(x,y)\bigr|^2.
		\label{eq:intensity_3d}
	\end{equation}
	
	Although the incident wave is purely TE-polarized ($\mathbf{E} \parallel \mathbf{y}$), the 3D mask (with a two-dimensionally periodic structure $\eps(x,y)$) generates cross-polarized scattering, so that the scattering amplitude $A_{x;m,n}^{(r)}$ is non-zero (the incident field contains no cross-polarized component, $A_{x;m,n}^{(i)} = 0$).
	
	The 3D latent function is:
	\begin{equation}
		f_{\mathrm{latent}}(x,y) = \mathrm{Re}\left(\sum_{m=-M}^{M} \sum_{n=-M}^{M} F_{m,n} \mathrm{e}^{-\mathrm{i}\kappa_x m x - \mathrm{i}\kappa_y n y}\right),
	\end{equation}
	where $F_{m,n}$ are the complex Fourier coefficients to be optimized (the two-dimensional generalization of $F_m$ in Eq.~\eqref{eq:latent}). The physical absorber density is obtained via the sigmoid mapping $\rho(x,y) = \sigma(\beta f_{\mathrm{latent}}) \in (0, 1)$, where $\beta$ is a steepness parameter increased during optimization to enforce binarization, and $\sigma$ is the sigmoid function defined in the 2D formulation. The permittivity is computed as $\eps(x,y) = \eps_{\mathrm{vac}} + \rho(x,y)(\eps_{\mathrm{abs}} - \eps_{\mathrm{vac}})$.

	\subsubsection{Parameters and Results of Optimization}
	
	Parameters: $\lambda = 11.2$~nm, incidence angle $6^\circ$ (TE), $L_x = L_y \approx 857$~nm, $L_x^{(\mathrm{w})} = L_y^{(\mathrm{w})} \approx 214$~nm, La absorber ($d_{\mathrm{abs}} = 60$~nm) on Ru/Be/Sr Bragg mirror, Fourier expansion parameter $M_{\mathrm{opt}} = M = 16$ (instead of the maximum propagating order $N = 19$), 500 epochs.
	
\begin{figure}[ht!]\centering
		\includegraphics[width=\textwidth]{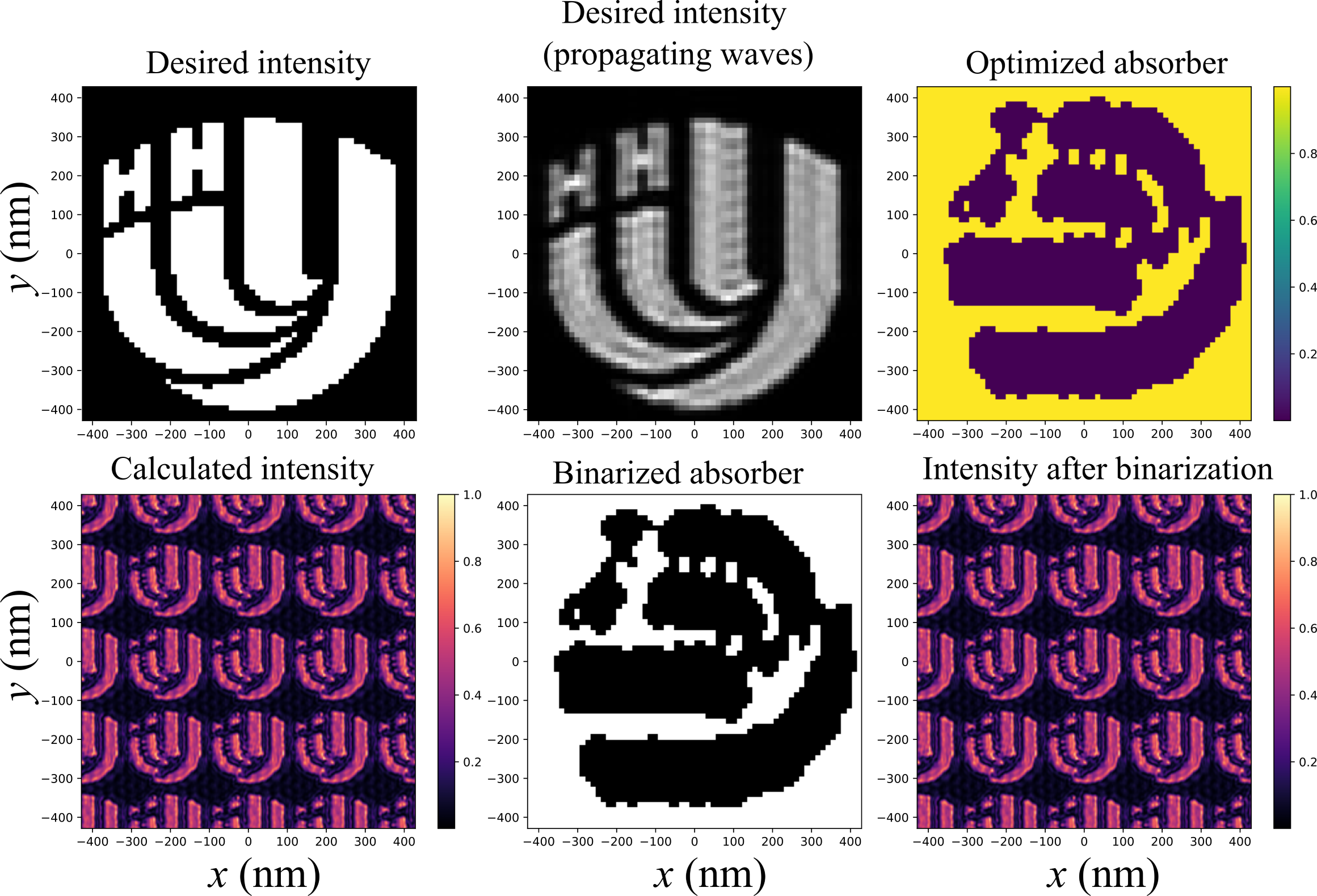}
		\caption{Optimization results for a 3D mask in the projection system ($\lambda=11.2$~nm, La absorber, $M_{\mathrm{opt}}=16$, $\mathrm{NA} \approx 0.84$, 500 epochs). Top row: target image (UNN logo), target projection onto propagating harmonics, optimized absorber density $\rho(x,y)$. Bottom row: synthesized aerial intensity on the wafer, binarized mask ($\rho > \langle\rho\rangle$), aerial intensity from the binarized mask. The mask period is $L_x = L_y \approx 857$~nm. The wafer period is $L_x^{(\mathrm{w})} \approx 214$~nm.}\label{fig:3d_mask_resultM16}
	\end{figure}

	Figure~\ref{fig:3d_mask_resultM16} presents the optimization results: target emblem (University of Nizhny Novgorod logo), bandlimited target projection, continuous optimized absorber density $\rho(x,y)$, synthesized aerial intensity, binarized mask, and aerial image from the binarized mask.
	
	The binarization penalty $\mathcal{L}_{\mathrm{bin}}$ drops from $0.25$ to $0.002$, and $\mathcal{L}_{\mathrm{shape}}$ decreases from $6.5\times10^{-8}$ to $2.3\times10^{-8}$. Truncating to $M = 16$ corresponds to $\mathrm{NA} = 0.84$.

	\section{Discussion: Advantages and Disadvantages of the Proposed Projection System}\label{discussion}
	
	\subsection{Advantages}
	\begin{enumerate}
		\item The array of $2N = 38$ planar facets independently maps each mask diffraction harmonic to the corresponding wafer harmonic up to $\mathrm{NA}_{\max} = 0.993$, a value unattainable in conventional multi-mirror EUV projection objectives with a limited number of reflections.
		
		\item For the periodic structures considered here, the common illuminated wafer region can have an area comparable to the illuminated mask area, apart from edge effects. The unit-cell period is reduced by four along each transverse direction, so more repeated cells fit within the illuminated area. This does not imply demagnification of an arbitrary finite mask pattern while preserving its image area.
		
		\item The finite facet apertures and angular selectivity of the Bragg coatings can act as spatial-frequency filters for individual channels. Suppression of unwanted angular components can reduce flare, but its effect on useful signal and aerial-image contrast requires a quantitative calculation with finite source bandwidth and angular extent, mask scattering, and facet acceptance.
		
		\item The two reflections retain approximately $50$--$60\%$ of the power leaving the mask in an accepted diffraction order. This channel transmission is not the fraction of the total mask-reflected power or source power delivered to a useful image region. Comparison values of about $12\%$ for six reflections and $2.8\%$ for ten reflections refer only to products of mirror reflectances.
		
		\item Each accepted harmonic undergoes the spatial-frequency transformation $m\kappa_x\to-4m\kappa_x$, corresponding to fourfold reduction and inversion of the periodic coordinate. Order-dependent complex reflection coefficients and aperture truncation modify the pattern, so exact reproduction of its shape is not implied by the linear frequency map alone.
		
		\item In Examples 1 and 2, all $N$ beams propagate strictly parallel to the $z$-axis after the first reflection, simplifying mutual alignment. In Example 3, beams propagate at deflection angles $(\theta_m^{(\mathrm{w})} - \theta_m)$ ranging from $\sim 2^\circ$ to $\sim 69^\circ$, and in Example 4 at tailored angles $\beta_m$.
		
		\item In Examples 1 and 2, tilt angles of the first mirror do not exceed $7.2^\circ$, rendering it nearly planar and amenable to precision machining with sub-micrometer accuracy. In Example 3, tilt angles reach $27.2^\circ$, and in Example 4, $33.4^\circ$.
		
		\item All facets are planar, so Bragg coatings can be laterally homogeneous across each facet, eliminating the complex graded-layer deposition required on aspheric surfaces in conventional systems~\cite{Kalden2025}.
		
		\item Separate facet pairs permit channel-specific coating optimization and can allow modular changes to the accepted numerical aperture and local replacement of degraded elements. Their practical benefit depends on alignment, phase control, and thermal and mechanical stability. Replacement without optical readjustment has not been demonstrated.
	\end{enumerate}
	
	\subsection{Disadvantages}
	\begin{enumerate}
		\item Changing $L_x$ or $L_y$ changes the diffraction angles and can require new facet positions, tilts, coatings, and channel counts. Feasibility depends on the available mechanical clearance, optical acceptance, phase tolerances, and source bandwidth. It is not established for arbitrary periods.
		
		\item Projections of the beams on the wafer grow as $L_m^{(\mathrm{w})} = a_m/\cos\theta_m^{(\mathrm{w})}$: values remain below $18$~mm for $m \le 16$, reach $28.7$~mm for $m=18$, and $82.1$~mm for $m=19$ (at $\theta_{19}^{(\mathrm{w})} = 83.2^\circ$, the projection is stretched by a factor of $8.5$). This is not related to diffractive beam spreading but is a geometric consequence of the grazing projection of a finite-width beam ($a_m \approx 10$~mm) onto the wafer plane. The effect can be mitigated by apodization or by sizing the apertures of the high-order facets.
		
		\item In Example 2, the second-mirror facets operate at grazing angles from $1.5^\circ$ to $41.6^\circ$. The facet length required for full beam interception reaches $382.5$~mm for $m=1$, whereas the corresponding lengths in Example 4 are approximately $10$~mm.
		
		\item In Example 4, the first-mirror facet centers span $\pm256.3$~mm, while the second-mirror centers reach $\pm1102.1$~mm. The complete transverse span is therefore approximately $2204.2$~mm using the rounded tabulated coordinates, before mechanical supports and facet extents are included. The total optical path length is approximately $3$~m.
		
		\item Preserving the energy balance across diffraction harmonics requires individual optimization of Bragg bilayer thicknesses for each first-mirror facet, covering incidence angles from $1.5^\circ$ to $47.8^\circ$.
		
		\item The optical path differences in Examples 1--3 introduce deterministic relative propagation phases, rather than destroying coherence under strictly monochromatic illumination. Example 4 removes these path-dependent phase differences. Position tolerances must be set by the allowable wavefront error, while a finite source bandwidth additionally requires temporal-coherence analysis.
		
		\item A normal surface displacement $\delta h$ produces an optical-path error of approximately $2\delta h\cos i$. For independent errors on two nearly normally reflecting facets, $\sigma_P\approx2\sqrt{2}\,\sigma_h$. As an illustrative wavefront criterion, $S\approx\exp[-(2\pi\sigma_P/\lambda)^2]\ge0.8$ at $\lambda=11.2$~nm gives $\sigma_h\lesssim0.30$~nm per facet. The actual surface and alignment tolerances must be derived from the aerial-image requirements and error correlations, especially across 2256 facets.
		
		\item The validity of treating diffraction orders as bounded collimated beams is analyzed in Appendix~\ref{appA}, confirming high precision for centimeter-scale EUV beams.
	\end{enumerate}

	\section{Conclusion}\label{conclusion}
	
	In this work, an all-reflective EUV projection concept has been proposed and investigated that uses two reflections per accepted channel between the mask and wafer, with each mirror composed of individual planar facets corresponding to each spatial diffraction order.
	
	The main conclusions are:
	\begin{enumerate}
		\item A two-mirror projection concept providing $4\times$ reduction of the periodic unit cell at $\mathrm{NA}_{\max}\approx 0.993$ is analyzed at $13.5$~nm and $11.2$~nm, with $L_x/\lambda$ kept fixed.
		\item A spatial geometry providing rigorous equalization of optical path lengths across all diffraction harmonics is established, removing order-dependent propagation phase shifts.
		\item Multi-parameter optimization of 30-bilayer Bragg mirrors (Mo/Si at $13.5$~nm and Ru/Be at $11.2$~nm) demonstrates channel throughput of $\sim 51$--$60\%$. Facets of the second mirror operate at quasi-normal incidence ($\le 3.93^\circ$) suggesting the possibility of a common coating as a separate approximation.
		\item The concept is generalized to 3D periodic masks. It has been shown that for the considered mask, the 2256-facet system with 1128 selected orders requires optimizing 132 radial coating pairs, or up to 264 distinct multilayer structures, due to rotational symmetry.
		\item Inverse lithography produces simulated aerial images with a central peak of approximately $5.4$~nm FWHM from a $4.6$-nm-wide target and with two $6$-nm-wide target lines separated by $6$~nm. The unused back-reflection order is suppressed in the 2D optimization, and the two peaks remain resolved for the tested defocus values up to $5$~nm. Resist printing and a full process window have not been evaluated.
		\item The validity of the geometrical-optics beam approximation for centimeter-scale EUV beams over $\sim 1$~m paths is proven (collimation length exceeds hundreds of meters, and edge blur is under a millimeter).
	\end{enumerate}
	
	The modular faceted approach offers a route to reducing the number of reflections in periodic-pattern projection while using planar reflecting surfaces. Its applicability to high-resolution lithography requires further assessment of source coherence and bandwidth, polarization, fabrication and alignment tolerances, and resist response. This faceted approach can removes the fundamental technological constraints of multi-mirror EUV optics associated with the number of reflections and complex aspherization, opening a promising route toward projection lithography systems of ultra-high resolution ($\mathrm{NA} \to 1$).

%

	\newpage
	\appendix
	
	\section{Diffractive Propagation of a Hard-Apertured Beam}\label{appA}
	
	\subsection{Problem Formulation}
	
	As in the main text, the time dependence $\exp(\mathrm{i}\omega t)$ is assumed and omitted. Let the initial field at $z = 0$ be non-zero only over an aperture of width $W$ along $x$:
	\begin{equation}
		\label{eq:initial}
		E(x, 0) = E_0\,\Pi\!\left(\frac{x}{W}\right) \mathrm{e}^{-\mathrm{i} k_x x},
		\qquad
		\Pi(u) = \begin{cases} 1, & |u| \le 1/2, \\ 0, & |u| > 1/2, \end{cases}
	\end{equation}
	where $k_x$ is the transverse wavevector component:
	\begin{equation}
		\label{eq:kvec}
		\mathbf{k} = (k_x, 0, k_z), \qquad
		k_x^2 + k_z^2 = k_0^2, \qquad
		k_0 = \frac{2\pi}{\lambda}.
	\end{equation}
	The propagation angle relative to the $z$-axis is $\alpha = \arcsin(k_x/k_0)$ ($\cos\alpha = k_z/k_0$).

	\subsection{Angular Spectrum}
	
	Expanding Eq.~\eqref{eq:initial} into plane waves:
	\begin{equation}
		\label{eq:angspec}
		E(x, z) = \int_{-\infty}^{\infty} A(q)\,
		\mathrm{e}^{-\mathrm{i} q x - \mathrm{i} \gamma(q) z}\,\frac{\mathrm{d}q}{2\pi},
		\qquad
		\gamma(q) = \left(k_0^2 - q^2\right)^{1/2},\quad \mathrm{Im}\,\gamma(q)\le0,
	\end{equation}
	with spectral amplitude:
	\begin{equation}
		\label{eq:spectrum}
		A(q) = E_0 W \sinc\!\left[\frac{(q - k_x) W}{2}\right],
		\qquad
		\sinc u \equiv \frac{\sin u}{u}.
	\end{equation}
	The spectral half-width to first zeros is $\Delta q = 2\pi/W$. The divergence half-angle is:
	\begin{equation}
		\label{eq:theta0}
		\theta_0 = \frac{\Delta q}{k_0 \cos\alpha}
		= \frac{\lambda}{W\cos\alpha}
		= \frac{\lambda k_0}{W k_z}
		= \frac{\lambda}{W_\perp},
	\end{equation}
	governed by the perpendicular beam width:
	\begin{equation}
		\label{eq:wperp}
		W_\perp = W\cos\alpha = \frac{W k_z}{k_0}.
	\end{equation}

	\subsection{Paraxial Approximation and Beam Coordinate System}
	
	Using the beam coordinates $(\xi,\zeta)$ defined below, write $E(x,z)=\exp(-\mathrm{i}k_0\zeta)U(\xi,\zeta)$. Neglecting $\partial^2U/\partial\zeta^2$ relative to $2\mathrm{i}k_0\partial U/\partial\zeta$ gives the paraxial envelope equation
	\begin{equation}
		\label{eq:paraxial}
		2 \mathrm{i} k_0 \frac{\partial U}{\partial\zeta} = \frac{\partial^2 U}{\partial\xi^2}.
	\end{equation}
	Equation~\eqref{eq:paraxial} describes diffraction along the beam axis. In laboratory coordinates, an envelope extracted using $\exp[-\mathrm{i}(k_xx+k_zz)]$ also has a transverse transport term. The spectral curvature is $\gamma^{\prime\prime}(k_x)=-k_0^2/k_z^3$.
	
	In the beam coordinate frame $(\xi, \zeta)$:
	\begin{equation}
		\label{eq:beamframe}
		\zeta = x\sin\alpha + z\cos\alpha \quad (\text{along the beam axis}), \qquad
		\xi   = x\cos\alpha - z\sin\alpha \quad (\text{transverse to the beam}).
	\end{equation}
	Along the central ray, a propagation distance $z$ in the laboratory frame corresponds to $\zeta = z/\cos\alpha = z k_0/k_z$ along the beam axis.
	
	To leading paraxial order, the oblique aperture is represented on the transverse beam plane by a top-hat envelope of width $W_\perp$. Its propagation is described by the Fresnel integral
	\begin{equation}
		\label{eq:fresnel}
		U(\xi, \zeta) = \sqrt{\frac{k_0}{-2\pi \mathrm{i} \zeta}}
		\int_{-W_\perp/2}^{W_\perp/2}
		U(\xi', 0)\,
		\exp\!\left[-\frac{\mathrm{i} k_0 (\xi - \xi')^2}{2\zeta}\right] \mathrm{d}\xi'.
	\end{equation}

	\subsection{Key Diffraction Scales}
	
	From Eq.~\eqref{eq:fresnel}, three characteristic scales emerge (where $L \equiv \zeta$ is propagation distance):
	
	\noindent\emph{1. Fresnel number ($N_F$) and collimation length ($L_F$):}
	\begin{equation}
		\label{eq:nf}
		N_F(L) = \frac{(W_\perp/2)^2}{\lambda L} = \frac{L_F}{L},
		\qquad
		L_F = \frac{W_\perp^2}{4\lambda} = \frac{W^2 k_z^2}{4 k_0^2 \lambda}.
	\end{equation}
	Near-field geometrical propagation holds for $N_F \gg 1$, while Fraunhofer diffraction occurs for $N_F \ll 1$.
	
	\noindent\emph{2. Near-Field Boundary Blur:} The edge transition scale is:
	\begin{equation}
		\label{eq:edge}
		\Delta \sim \sqrt{\lambda L} = \sqrt{\frac{2\pi L}{k_0}},
	\end{equation}
	with the $10\%$--$90\%$ intensity transition width given by:
	\begin{equation}
		\label{eq:edge1090}
		\Delta_{10\text{--}90\%} \approx 0.829\,\sqrt{\lambda L}.
	\end{equation}
	
	\noindent\emph{3. Far-Field Divergence:} The angular intensity profile is:
	\begin{equation}
		\label{eq:sinc2}
		I(\theta) = I_0 \sinc^2\!\left(\frac{\pi W_\perp \theta}{\lambda}\right),
	\end{equation}
	with a main-lobe angular width between the first zeros of $2\theta_0 = 2\lambda/W_\perp$, a full width at half maximum of $\mathrm{FWHM} \approx 0.886\,\lambda/W_\perp$, and a first sidelobe of $\approx 4.7\%$ of the maximum. The beam size expands as $D(L) \approx 2\theta_0 L$.

	\subsection{Near-Field Boundary Structure}
	
	When $\sqrt{\lambda L} \ll W_\perp$ ($N_F \gg 1$), opposite aperture edges diffract independently, matching semi-infinite knife-edge diffraction:
	\begin{equation}
		\label{eq:cornu}
		I(v) = \frac{I_0}{2}\left\{\left[C(v) + \tfrac12\right]^2
		+ \left[S(v) + \tfrac12\right]^2\right\},
		\qquad
		v = (\xi_{\mathrm{edge}}-\xi)\sqrt{\frac{2}{\lambda L}},
	\end{equation}
	where $C(v)$ and $S(v)$ are Fresnel integrals and $\xi_{\mathrm{edge}}$ is the right aperture edge, so $v>0$ points into the illuminated region. Characteristics include:
	\begin{itemize}
		\item At the shadow boundary, $I(0) = I_0/4$.
		\item The first diffraction peak reaches $I \approx 1.37\,I_0$ at $v \approx 1.22$.
		\item In the illuminated region, decaying Fresnel fringes occur with period $\sim \lambda L / |\xi - \xi_{\mathrm{edge}}|$.
		\item Field decays monotonically into the shadow region.
	\end{itemize}
	
	The number of Fresnel oscillations across the beam width is of the order of $N_F$. As $N_F$ decreases to $\sim 10$, the oscillations from the opposite edges begin to interfere with each other, distorting the central part of the plateau. At $N_F \sim 1$ the profile smooths out and transitions into the far-field distribution~\eqref{eq:sinc2}.

	\subsection{Numerical Example: $\lambda = 11.2$~nm, $W = 1$~cm, $k_x = k_z$}
	
	For $k_x = k_z = k_0/\sqrt{2}$ ($\alpha = 45^\circ$):
	\begin{center}
		\begin{tabular}{ll}
			\toprule
			Parameter & Value \\
			\midrule
			$k_0 = 2\pi/\lambda$ & $5.61 \times 10^{8}$~m$^{-1}$ \\
			$k_x = k_z$ & $3.97 \times 10^{8}$~m$^{-1}$ \\
			$W_\perp = W/\sqrt{2}$ & $7.07$~mm \\
			$\theta_0 = \lambda/W_\perp$ & $1.58 \times 10^{-6}$~rad $\approx 0.33''$ \\
			$L_F = W_\perp^2/(4\lambda)$ & $1.1 \times 10^{3}$~m \\
			\bottomrule
		\end{tabular}
	\end{center}
	
\begin{figure}[ht!]\centering
		\includegraphics[width=0.5\textwidth,keepaspectratio]{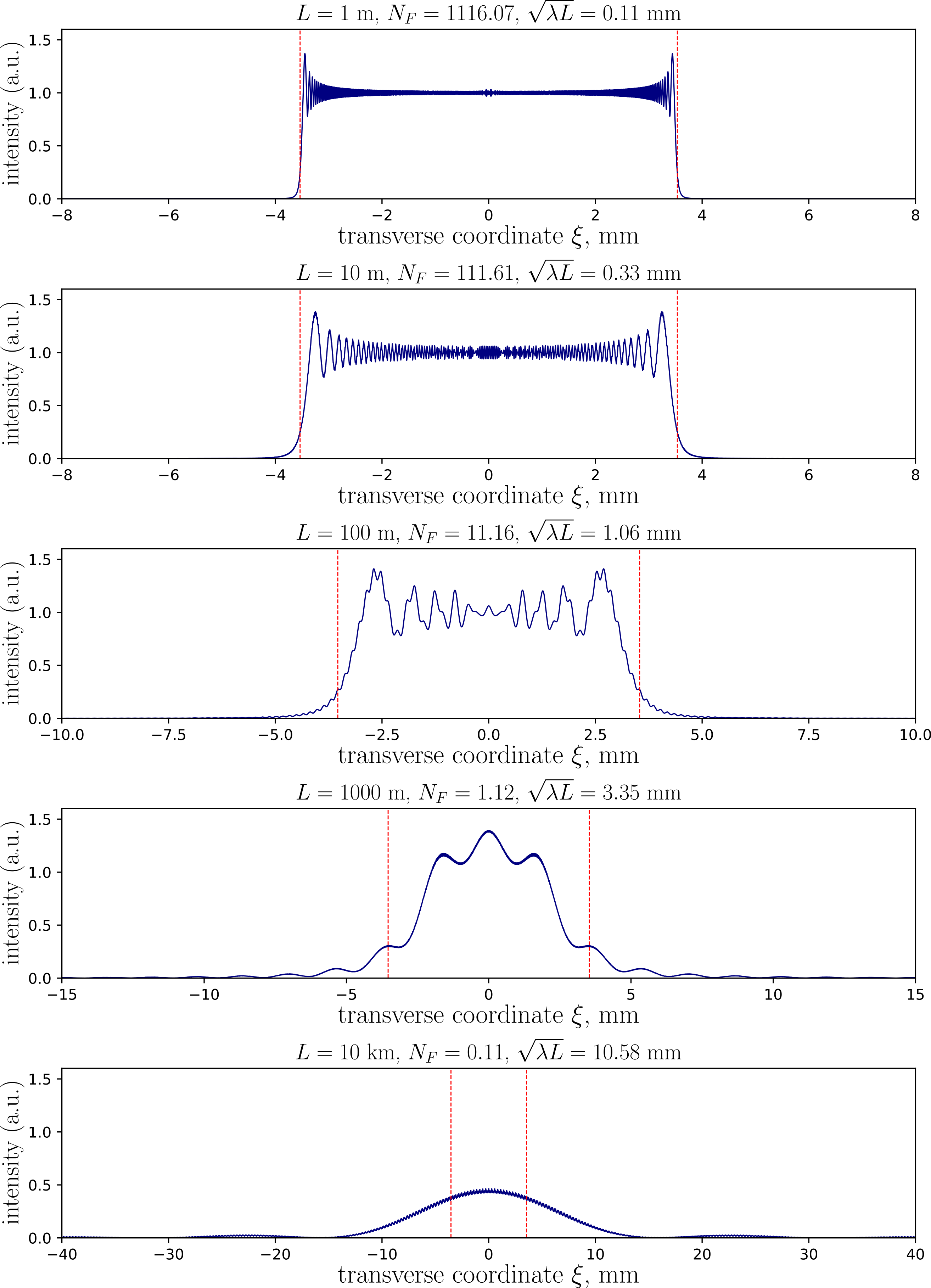}
		\caption{Transverse intensity profile evolution versus distance (angular spectrum calculation). Dashed lines: geometrical shadow boundaries.}\label{fig:profiles}
	\end{figure}

\begin{figure}[ht!]\centering
		\includegraphics[width=\textwidth]{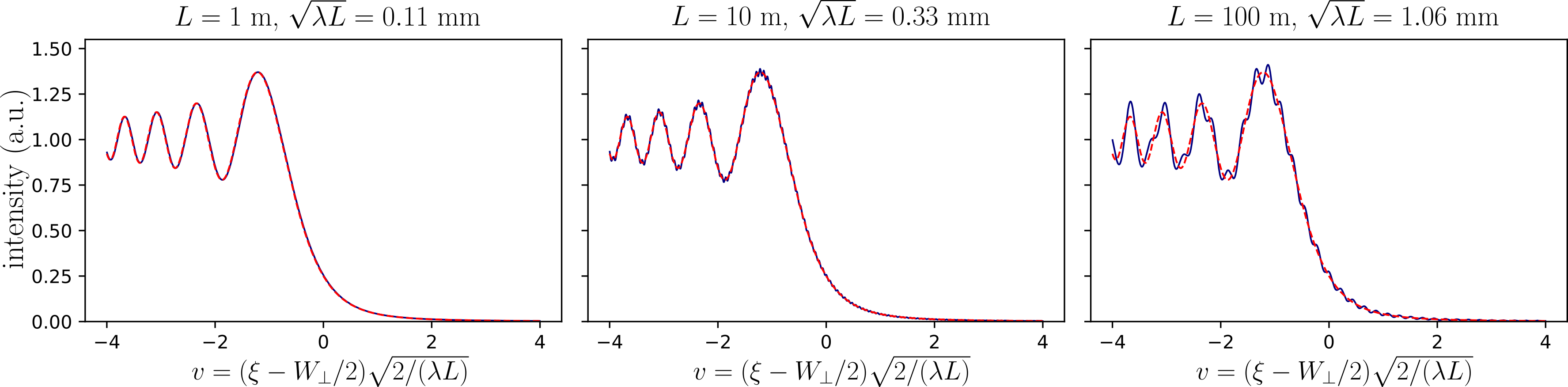}
		\caption{Right-edge boundary diffraction profile. The solid line shows the slit calculation. The dashed line shows the semi-infinite knife-edge analytical model, Eq.~\eqref{eq:cornu}. Small oscillations at $L = 100$~m arise from interference with the wave diffracted by the opposite edge.}\label{fig:edge}
	\end{figure}

\begin{figure}[ht!]\centering
		\includegraphics[width=0.5\textwidth,keepaspectratio]{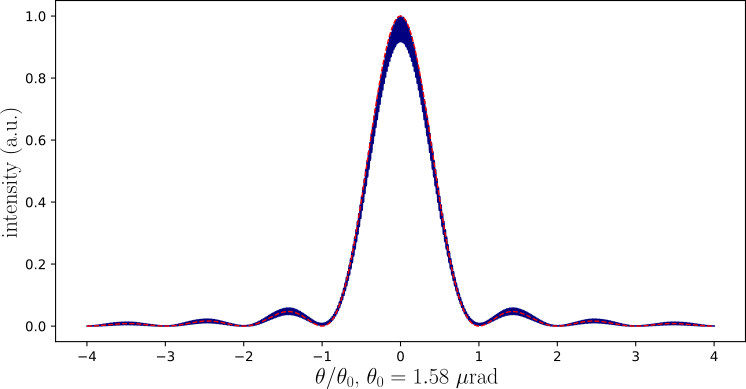}
		\caption{Far-field intensity distribution ($L = 10$~km, $N_F = 0.11$): the solid line shows the numerical result in normalized angular coordinate $\theta/\theta_0$. The dashed line shows the theoretical $\sinc^2(\pi W_\perp \theta/\lambda)$.}\label{fig:farfield}
	\end{figure}

	Table~\ref{tab:results} and Figs.~\ref{fig:profiles}--\ref{fig:farfield} illustrate beam evolution across distances.
	
	\begin{table}[ht]
		\centering
		\caption{Diffractive beam evolution with propagation distance ($\lambda = 11.2$~nm, $W_\perp = 7.07$~mm). The $10\%$--$90\%$ widths at $L\le100$~m use the independent-edge approximation. The $L=1000$~m value is a numerical finite-slit estimate, where the two edge fields overlap and Eq.~\eqref{eq:edge1090} is not applicable.}
		\label{tab:results}
		\begin{tabular}{rrrrl}
			\toprule
			$L$, m & $N_F$ & $\sqrt{\lambda L}$ (mm) & $\Delta_{10\text{--}90\%}$ (mm) & Propagation Regime \\
			\midrule
			1     & 1116 & 0.11 & 0.09 & Geometrical shadow \\
			10    & 112  & 0.33 & 0.28 & Near field \\
			100   & 11   & 1.06 & 0.88 & Edge-ripple overlap \\
			1000  & 1.1  & 3.35 & 2.30 & Intermediate zone \\
			10000 & 0.11 & 10.6 & ---  & Fraunhofer far field ($\mathrm{FWHM} = 14$~mm) \\
			\bottomrule
		\end{tabular}
	\end{table}
	
	\subsection{Conclusions of the Diffraction Analysis}
	\begin{enumerate}
		\item A centimeter-aperture EUV beam ($W_\perp \approx 7$~mm) at $\lambda = 11.2$~nm exhibits high collimation stability: $L_F \approx 1.1$~km and divergence $\theta_0 \approx 1.6~\mu$rad. Over typical optical paths ($L \sim 1$--$3$~m), the beam profile is practically indistinguishable from the geometrical projection of the mask.
		\item Edge blur obeys $\Delta \sim \sqrt{\lambda L}$, amounting to only $0.09$~mm at $1$~m and $0.28$~mm at $10$~m. Near the boundaries, Fresnel fringes form with a local peak reaching $1.37\,I_0$, decaying monotonically into the shadow.
		\item At distances $L \gtrsim L_F$, sharp edges blur and the beam transitions into a $\sinc^2$ distribution with sidelobe levels of $\approx 4.7\%$. The beam diameter grows linearly ($D \approx 2\theta_0 L$).
		\item Hard aperture truncation generates edge ripples. If suppression is required, apodizing elements with smoothed edge transmission/reflection can be incorporated.
	\end{enumerate}

	\section{Nomenclature}\label{appB}
	
	This table summarizes the main abbreviations used in the paper.
		\begin{table}[h!]
			\centering
			\caption{Nomenclature}\label{tableAppB}
				\begin{tabular}{ll}
					Notation & Description  \\
					\hline
					CAD & Computer-aided design \\
					CD & Critical dimension \\
					DoF & Depth of focus \\
					EUV & Extreme ultraviolet \\
					FWHM & Full width at half maximum \\
					GO & Geometrical optics \\
					ILT & Inverse lithography technology \\
					L-BFGS-B & Bound-constrained limited-memory BFGS method \\
					Mo/Si & Molybdenum/silicon multilayer Bragg coating \\
					MSE & Mean squared error \\
					NA & Numerical aperture \\
					OAI & Off-axis illumination \\
					RCWA & Rigorous coupled-wave analysis \\
					RMS & Root mean square \\
					Ru/Be & Ruthenium/beryllium multilayer Bragg coating \\
					Ru/Be/Sr & Ruthenium/beryllium/strontium multilayer Bragg coating \\
					TE & Transverse electric polarization \\
					TM & Transverse magnetic polarization \\
					TMM & Transfer-matrix method \\
					HF & High-frequency penalty \\
					UNN & University of Nizhny Novgorod \\
				\end{tabular}
	\end{table}
	
	\newpage
	\bibliographystyle{IEEEtran}
	\bibliography{Eskin_EUV_mirror}
	
\end{document}